\documentclass[10pt,a4paper]{article} 
\usepackage[utf8]{inputenc}
\usepackage{amsmath}
\usepackage[unicode,colorlinks=true,linkcolor=blue,citecolor=blue]{hyperref}
\def\equationautorefname~#1\null{(#1)\null}
\usepackage[all]{hypcap}
\usepackage{textcomp}
\usepackage{amsfonts}
\usepackage{mathrsfs}
\usepackage{lettrine}
\usepackage{amssymb}
\usepackage{graphicx}
\usepackage{graphicx}
\usepackage{hyperref}
\usepackage[superscript]{cite}
\usepackage{vmargin}
\usepackage{eurosym}
\setmarginsrb{20mm}{25mm}{20mm}{30mm}{10pt}{0mm}{0pt}{20mm}
\makeindex
\usepackage{array}
\usepackage{bbm}
\usepackage{multicol}
\usepackage{bm}
\usepackage{lipsum}
\usepackage{dcolumn}
\usepackage{braket}
\usepackage{stmaryrd}
\usepackage{color}
\usepackage{rotating}
\usepackage{longtable}
\usepackage{soul}
\nonstopmode
\usepackage{tocloft}
\DeclareMathAlphabet{\mathpzc}{OT1}{pzc}{m}{it}
\let\oldref\ref
\renewcommand{\ref}[1]{(\oldref{#1})}

\begin{document}

\title{\vspace*{-2cm} Charge separation: From the topology of molecular electronic transitions to the dye/semiconductor interfacial energetics and kinetics}
\date{}
\date{}

\maketitle

\vspace*{-1.4cm}

\noindent \begin{center}
Thibaud Etienne$^{a,\dag}$ and Mariachiara Pastore$^{b,c,\ddag}$
\end{center}

$\;$

\noindent $^a$ Institut Charles Gerhardt -- CNRS and Université de Montpellier, Place Eugène Bataillon -- 34095 Montpellier, France \\
\noindent $^b$ Université de Lorraine -- Nancy, Laboratoire de Physique et Chimie Théoriques, Boulevard des Aiguillettes, 54506 Vandoeuvre-lès-Nancy, France \\
\noindent $^c$ CNRS -- Laboratoire de Physique et Chimie Théoriques, Boulevard des Aiguillettes, 54506 Vandoeuvre-lès-Nancy, France

$\;$

\noindent $^\dag$ thibaud.etienne@umontpellier.fr \\
\noindent $^\ddag$ mariachiara.pastore@univ-lorraine.fr 

$$$$

\noindent\textbf{Abstract}
Charge separation properties, that is the ability of a chromophore, or a chromophore/semiconductor interface, to separate charges upon light absorption, are crucial characteristics for an efficient photovoltaic device. Starting from this concept, we devote the first part of this book chapter to the topological analysis of molecular electronic transitions induced by photon capture. Such analysis can be either qualitative or quantitative, and is presented here in the framework of the reduced density matrix theory applied to single-reference, multiconfigurational excited states. The qualitative strategies are separated into density-based and wave function-based approaches, while the quantitative methods reported here for analysing the photoinduced charge transfer nature are either fragment-based, global or statistical. In the second part of this chapter we extend the analysis to dye-sensitized metal oxide surface models, discussing interfacial charge separation, energetics and electron injection kinetics from the dye excited state to the semiconductor conduction band states.

$\;$


\section{Introduction}

In the context of solar energy exploitation, dye-sensitized solar cells (DSCs) \cite{Graetzel-2003, Graetzel-2009, Hagfeldt-2010} 
and dye-sensitized photoelectrosynthetic cells (DSPECs)\cite{Alibabaei-2013, Walter-2010, Youngblood-2009, Yu-2015} offer the promise of cost effective 
sunlight conversion and storage, respectively. Since the seminal paper by O'Regan and Gr{\"a}tzel in 1991,\cite{oregan-1991} hybrid/organic 
photovoltaic devices have attracted significant research interest, with the recent launch of the first commercial product. 
The heart of a \textit{n}-type DSC is the photoanode, where a mesoporous oxide layer, usually TiO$_2$ or ZnO, is sensitized by a monolayer of dyes, chemically bound to the semiconductor nanoparticles. Upon solar irradiation, the adsorbed sensitizers are able to inject the photoexcited electrons in the manifold of the conduction band (CB) states of the 
semiconductor, typically at the femtosecond time scale. Injected electrons travel through the mesoporous film and are collected by the conductive layer of the photoanode electrode, while the oxidized dye is rapidly reduced by the electron mediator donor in solution or by the hole transporting material (HTM) in solid-state devices. The collected electrons flow in the external circuit, producing a photocurrent, and reach the counter-electrode, where the circuit is closed by the reduction of the electron mediator acceptor.
Alternatively, in a \textit{p}-type DSC, the photoactive electrode is a dye-functionalized nanostructured wide band gap \textit{p}-type semiconductor 
(usually NiO).\cite{He-1999, Borgstrom-2005, Odobel-2012, Brennaman-2016} In this case the functioning mechanism is inverse, with hole injection from the photoexcited dye to the 
\textit{p}-type semiconductor valence band (VB) and the concomitant electron transfer to the redox mediator and final electron collection at the anode.  

Therefore, the key-phenomena at the center of the electron current photogeneration in DSCs are light-matter interaction and the subsequent intramolecular and interfacial charge separation.\cite{PastoreTopCurrChem2014,PCCP_Iron} Indeed, when a molecule catches the light, it sees its electronic structure totally reorganized. While approaching theoretically the electronic structure of complex molecular systems is already quite a challenge,\cite{PastoreJPhysChemC2013,RoncaJChemTheoryComput2014} providing a reliable prediction about its time evolution due to the photoactivity of the molecule constitutes an arduous task.\cite{LabatAccChemRes2012,jacquemin_accurate_2009} The keys to a proper description of the light-induced electronic cloud polarization are contained into the so-called exciton wave function, and several tools exist for unraveling the nature of molecular excited states, based the exciton analysis.\cite{plasser_analysis_2012,plasser_new_2014,plasser_newbis_2014,bappler_exciton_2014,mewes2015communication,plasser2015statistical,wenzel_physical_2016,
plasser2016entanglement,mai_quantitative_2018} Using the reduced density matrix theory for extracting and analyzing this information has already been proved to be of interest to theoretical and computational communities,\cite{etienne2015transition,etienne_theoretical_2018} but also in more applied research fields,\cite{dedeoglu_detection_2014-3,etienne_qm/mm_fluo,etienne_fluorene-imidazole_2016,ho2017vibronic} and in particular in the DSCs community.\cite{etienne_iron,etienne_all-organic_2014,duchanois_heteroleptic_2014,duchanois_iron,Liu-2016} Indeed, some strategies have already been developed for providing qualitative and quantitative insights into the topology of molecular electronic transitions with the projection of the exciton wave function in a finite-dimensional vector space (the transition density matrix\cite{martin_natural_2003,batista_et_martin_natural_2004,dreuw_single-reference_2005,mayer_using_2007,surjan_natural_2007,furche_density_2001,luzanov_electron_2010,luzanov_interpretation_1980,plasser_new_2014,YonghuiLi_2001,tretiak_density_2002,wu2008exciton,tretiak2005exciton,etienne2015transition}) as a starting point.\cite{plasser_new_2014,etienne_theoretical_2018} The most common approaches consist in constructing two one-particle wave functions (occupied/virtual Natural Transition Orbitals\cite{etienne2015transition,martin_natural_2003
,batista_et_martin_natural_2004,dreuw_single-reference_2005,mayer_using_2007,surjan_natural_2007}) or charge density functions (detachment/attachment densities\cite{HeadGordonJPhysChem1995,dreuw_single-reference_2005,pastore2017unveiling}) which allow for a straightforward perspective of the photo-induced electronic structure reorganization.\cite{plasser_new_2014,etienne_theoretical_2018} A quantitative probing of the exciton locality, nature and structure, has also been derived.\cite{luzanov_interpretation_1980,peach_excitation_2008,luzanov_electron_2010,le_bahers_qualitative_2011,garcia_evaluating_2013,GuidoJChemTheoryComput2013,etienne_toward_2014,etienne_new_2014,guido_effective_2014,etienne_probing_2015,
wenzel_physical_2016,savarese_metrics_2017,mai_quantitative_2018,etienne_theoretical_2018} 

The first part of this book chapter is devoted to the description of the mathematical details of the strategies established for assessing the locality and nature of molecular electronic transitions, as the knowledge about the potentiality of a molecule to separate charge in space is obviously extremely helpful when designing new dyes for solar cells applications. The second part discusses how the intramolecular charge separation (electronic structure and excited state properties) of the dye affects the interfacial energetics and overall charge generation characteristics of dye-sensitized photoelectrode models. We will start by investigating the ``perturbation'' of the dye and semiconductor energy levels due to the dye chemisorption on 
the TiO$_2$\cite{RoncaEnergyEnvironSci2013,PastoreTopCurrChem2014} and NiO\cite{Piccinin-2017} surfaces to then move to the modelling of the dye and semiconductor electronic coupling and charge injection as a function of the dye molecular structure and substrate protonation.\cite{RoncaJPhysChemC2014Prot}

\section{Topology of molecular electronic transitions}
In this section we will revisit various qualitative and quantitative tools for analyzing the charge displacement occurring upon light absorption, {i.e.}, the electronic structure reorganization induced by the capture of a photon. While qualitative analyses provide visualization tools for assessing the nature of a charge transfer, the quantitative approaches consist of numbers that can be regarded as a collection of data for numerical analyses of the electronic transition character. Among the different approaches considered in this chapter, we can split the qualitative strategies into two parts: Those involving the analysis of charge density functions constructed from one-particle reduced density matrices of different nature, and those involving the analysis or the production of wave functions. For the quantitative counterpart, one can also consider two sets of mathematical objects: The fragment-based and the global strategies. While the former will provide charge transfer or excitation numbers that will be system-dependent (more exactly, they will depend on how the whole electronic system is divided into fragments according to the molecular structure), the latter produce numbers that represent the whole molecular system and its electronic structure. They are said to be ``systematic" due to the fact that they target the entire, unfragmented system. 

The choice of density or wave functions for the qualitative analysis of charge transfers can also be guided by the choice of the quantitative analysis to be used afterward, since for instance some quantitative insights can be derived from the singular values corresponding to natural transition orbitals (\textit{vide infra}) while the hole/particle spatial overlap integral or the charge transfer integral can be computed more directly from detachment/attachment density functions (\textit{vide infra}).

 Note that for the sake of simplicity, through this section we will only depict the analysis of the outcome of calculations of electronic transitions between a single-reference electronic ground state and an excited state constructed as a linear combination of singly-excited Slater determinants (unless the contrary is explicitly stated). 
 
 Note finally that most of the features reported in this section can be found in software packages freely available on the web\cite{CTpackage,theodore,nancyex,taeles}
\subsection{Background - State density matrix and charge density}
As briefly recalled in Refs.\cite{etienne2015transition,etienne_theoretical_2018} to a quantum electronic state $\ket{\psi _m}$, projected in a spin-spatial space $\{ \textbf{x} \}$ where \textbf{x} collects spatial (\textbf{r}) and spin ($\sigma$) coordinates of the $N$ electrons of a molecular system, i.e.,
\begin{equation}
\ket{\psi _m} \leftrightarrow \psi _m (\textbf{x}) = \braket{\textbf{x}|\psi _m},
\end{equation}
corresponds a reduced one-particle density matrix kernel $\tilde{\gamma}_m$
\begin{align}
\tilde{{{\gamma}}}_m (\textbf{r}_1;\textbf{r}_1') = N \!\!\sum _{\sigma _1 = \alpha , \beta} \int \! d\textbf{x}_2 \ldots \int \! d\textbf{x}_N \; \psi _m (\textbf{r}_1, \sigma _1 ,..., \textbf{x}_N)  \, \psi _m ^* (\textbf{r}_1', \sigma _1, ... ,\textbf{x}_N).
\end{align}
When the wave function is expressed in the basis of $L$ one-particle functions (spinorbitals - SOs) $\{\varphi\}$, with $N$ of them being occupied, and $L-N$ unoccupied (virtual), $\tilde{{{\gamma}}}_m$ can be rewritten
\begin{equation}
\tilde{{{\gamma}}}_m (\textbf{r}_1;\textbf{r}_1') = \sum _{r,s=1} ^L  \varphi _r(\textbf{r}_1) \, (\bm{\gamma}_m)_{rs} \, \varphi _s ^* (\textbf{r}_1') \qquad ; \qquad (\bm{\gamma}_m)_{rs} = \braket{\psi _m | \hat{r}^\dag \hat{s}|\psi _m} \label{eq:tildegamma}
\end{equation}
where $\bm{\gamma}_m$ is simply termed one-particle reduced density matrix (1-RDM). The $\hat{r}^\dag$ and $\hat{s}$ operators are respectively the creation and annihilation operators from the second quantization. From the expression of $\tilde{{{\gamma}}}_m$ we see that the one-particle charge density (1-CD) for the quantum state $\ket{\psi _m}$ is the trace of $\tilde{{{\gamma}}}_m$
\begin{equation}
\tilde{\gamma}_m (\textbf{r}_1;\textbf{r}_1') = n_m(\textbf{r}_1) \; \forall \textbf{r}_1 = \textbf{r}_1'.
\end{equation}
We therefore see that for a given $\{\varphi\}$ basis, there is an unequivocal relationship between any 1-RDM and the corresponding 1-CD
\begin{equation}
\bm{\gamma} \overset{\mathbb{R}^3}{\longrightarrow} \, n(\textbf{r}_1) = \sum _{r,s=1}^L \varphi _r (\textbf{r}_1) (\bm{\gamma})_{rs}\varphi ^* _s (\textbf{r}_1). \label{eq:1rdm1cd}
\end{equation} 
Since the outcome of the post-processing of electronic excited states quantum chemical calculations (or the outcome of the calculations themselves) is often a 1-RDM, this relationship will often be used for projecting a density matrix into the Euclidean space for visualizing changes in the charge density. Such type of analysis will be the basis for several qualitative analyses of photoinduced electronic structure reorganization, as introduced in the next paragraph.
\subsection{Qualitative analysis of photoinduced electronic structure reorganization}
As previously stated, we will first expose the direct use or the transformation of state density matrices for visualizing the photoinduced charge density rearrangement using one-particle density functions (charge densities or wave functions). We will first expose qualitative topological analyses involving 1-CD functions. Note that from now on, when two position coordinate vectors, $\textbf{r}_1$ and $\textbf{r}_1'$, will be present in the expression of any function, these coordinates will respectively point to the hole and the particle generated by photon capture, similar to $\textbf{r}$ and $\textbf{r}'$ or $\textbf{r}_h$ and $\textbf{r}_e$ that one can meet in the literature.\cite{plasser_new_2014}
\subsubsection{Density-based strategies} 
As an outcome of an excited state quantum-chemical calculation, one obtains the 1-RDM of the ground ($\bm{\gamma} _0$) and the $x^{\mathrm{th}}$ excited state ($\bm{\gamma} _x$), together with the so-called one-particle transition density matrix (1-TDM, $\bm{\gamma}^{0x}$). These three matrices are the result of the projection of a reduced one-particle (transition) density matrix kernel into a finite SO space $\{\varphi\}$ (\textit{vide supra}). 
 \\ $\;$ \\ \newpage \noindent \textit{State densities} \\ $\;$ \\
According to equation \ref{eq:1rdm1cd}, one can obtain the projection of  $\bm{\gamma} _0$ and $\bm{\gamma} _x$ into the direct (Euclidean) space. Due to the linearity of such projection, it is also possible to unequivocally derive the 1-CD corresponding to the net change in the electron density between the excited and ground state from the difference of the state density matrices:
\begin{equation}
\bm{\gamma}_x - \bm{\gamma}_0 = \bm{\gamma}^{\Delta} \; \overset{\mathbb{R}^3}{\longrightarrow} \; n _{\Delta}(\textbf{r}_1) = n_x(\textbf{r}_1) - n_{0}(\textbf{r}_1).
\end{equation}
The so-called difference density, i.e., the difference between the 1-CD of the excited and ground states, can be used as a visualization tool to qualitatively locate the zones of space where the fluctuation of charge density is negative or positive. Two 1-CD functions ($n_\pm (\textbf{r}_1)$) can then be written, based on the sign of the difference density everywhere in $\mathbb{R}^3$, as\cite{le_bahers_qualitative_2011}
\begin{equation}
\dfrac{1}{2}\left\lbrace\sqrt{n^2 _\Delta(\textbf{r}_1)} \pm n_\Delta (\textbf{r}_1)\right\rbrace =  n_{\pm}(\textbf{r}_1) \; \in \; \mathbb{R}^+
\end{equation}
where we see that while the difference density can take positive or negative values, $n_+$ and $n_-$ are both positive functions where they are not vanishing. One can therefore separate the positive and the absolute value of the negative contributions to the difference density by using the $n_\pm$ function. Note that $n_+$ and $n_-$ are two 1-CD functions with strictly no spatial overlap.
\\ $\;$ \\ \textit{Charge displacement analysis} \\ $\;$ \\
It is possible, through the so-called Excited State Charge Displacement (ESCD) analysis,\cite{ronca_charge-displacement_2014,ronca_density_2014} to obtain a function along one coordinate (for example, $z$) which gives an idea of the charge displacement profile related to the transition between the ground and the excited states, along a molecular axis:
\begin{equation} 
\label{eqn:deltaq0x}
\Delta q^{0x}(z)=\displaystyle
\int_{-\infty}^{z} dz'
\int_{-\infty}^{+\infty} dx
\int_{-\infty}^{+\infty} dy \;
n_\Delta (x,y,z').  
\end{equation}
This type of analysis is quite general, and will also be used in another context (\textit{vide infra}) than the passage from ground to excited states.
\\ $\;$ \\ \textit{Detachment/attachment densities}\\ $\;$ \\
While the $n_\pm$ functions correspond to the net increase/decrease of charge density through space, one can also be interested in visualizing the hole and the particle density through two 1-CD functions, termed ``detachment" and ``attachment" densities.\cite{HeadGordonJPhysChem1995,dreuw_single-reference_2005,pastore2017unveiling} For constructing these two functions, one has first to diagonalize the difference density matrix and, as for the $n_\pm$ function in the Euclidean space, to use a $\textbf{k}_\pm$ matrix function in order to split the eigenvalues into two matrices according to their sign:
\begin{equation}
\exists \, (\textbf{M} , \textbf{k}_{\pm}) \; | \; \textbf{M}^\dag \textcolor{black}{\bm{\gamma}^{\Delta}} \textbf{M} = \textbf{m} \; , \; \textbf{k}_{\pm} = \dfrac{1}{2}\left\lbrace \sqrt{\textbf{m}^2} \pm \textbf{m}\right\rbrace. \label{eq:MdagM}
\end{equation}
Backtransforming the resulting diagonal matrices returns the detachment ($\bm{\gamma}^d$) and attachment ($\bm{\gamma}^a$) 1-RDM, that can be projected into the direct space (see equation \ref{eq:1rdm1cd})
\begin{equation}
\textbf{M} \textbf{k}_- \textbf{M}^\dag = \bm{\gamma}^d \; \overset{\mathbb{R}^3}{\longrightarrow} \; n _{d}(\textbf{r}_1) \qquad ; \qquad \textbf{M} \textbf{k}_+ \textbf{M}^\dag = \bm{\gamma}^a \; \overset{\mathbb{R}^3}{\longrightarrow} \; n _{a}(\textbf{r}_1).
\end{equation}
A quick look at these two 1-CD functions provides a clear idea of the location of the depletion/increment zones of electron density, generated by photon absorption (or emission). Note that unlike the $n_\pm$ functions, the detachment/attachment densities can overlap (see the top of Figure 2 in Ref.\cite{etienne_theoretical_2018}). This spatial overlap can even be quantified (\textit{vide infra}) and constitutes a measure of the exciton (i.e., the electron/hole pair) locality.
\\ $\;$ \\\textit{Transition densities} \\ $\;$ \\
As in the case of the 1-RDM kernel, one can write a one-particle reduced \textit{transition} density matrix kernel,\cite{plasser_new_2014,etienne2015transition,etienne_theoretical_2018} where we find that the product of a state wave function by its complex conjugate is replaced by the product of a state wave function by the complex conjugate of the wave function corresponding to another quantum electronic state:
\begin{align}
\tilde{\gamma}^{0x}(\textbf{r}_1;\textbf{r}_1') &= N \!\!\sum _{\sigma _1 = \alpha , \beta} \int \! d\textbf{x}_2 \ldots \int \! d\textbf{x}_N \; \psi _0 (\textbf{r}_1, \sigma _1 ,..., \textbf{x}_N)  \, \psi _x ^* (\textbf{r}_1', \sigma _1, ... ,\textbf{x}_N) \\ &= z_x^{-1/2}\sum _{r,s=1}^L   \varphi _r (\textbf{r}_1) (\bm{\gamma}^{0x})_{rs} \varphi _s ^* (\textbf{r}_1').
\end{align}
where $z_x^{-1/2}$ is a normalization factor (\textit{vide infra}). Here, we see that the couple of coordinates $(\textbf{r}_1$ and $ \textbf{r}_1')$ is taken so that the hole $(\textbf{r}_1)$ originates from the ground state and the particle $(\textbf{r}_1')$ belongs to the excited electronic state. This convention ($\textbf{r}_1$ coordinates for the hole, $\textbf{r}_1'$ for the particle) will be used through this book chapter. As in the case of state densities, the one-particle transition density function can be deduced from the kernel expression, and there is an unequivocal relation between the 1-TDM $z_x^{-1/2}\bm{\gamma}^{0x}$ and the transition density
\begin{equation}
\tilde{\gamma}^{0x}(\textbf{r}_1;\textbf{r}_1') = n _{0x}(\textbf{r}_1) \; \forall \textbf{r}_1 = \textbf{r}_1' \qquad \Rightarrow \qquad z_x^{-1/2}\bm{\gamma}^{0x}  \; \overset{\mathbb{R}^3}{\longrightarrow} \; n _{0x}(\textbf{r}_1)
\end{equation}
Considering the 1-TDM kernel as the exciton wave function, one can use its trace to visualize the transition density and use it for evaluating transition properties.
\subsubsection{Wave function-based strategies} 
Another possibility for visualizing the changes in the electronic structure of a molecule due to the capture of a photon is the use of one-particle wave functions (orbitals). Since the 1-TDM is given in the space of spinorbitals, one can interpret it as a compilation of the pondering coefficients describing an excited state as a superposition of single excitations from occupied to virtual spinorbitals, as explained below.
\\ $\;$ \\ \textit{Spinorbitals (SOs) analysis} \\ $\;$ \\
If an excited state $\ket{\psi _x}$ can be written as a linear combination of singly-excited Slater determinants
\begin{eqnarray} 
\ket{\psi _x} = \sum _{i=1}^N \sum _{a=N+1}^{L} z_x^{-1/2}(\bm{\gamma}^{0x})_{ia} \ket{{\psi _{\textcolor{black}{i}} ^{\textcolor{black}{a}}}} \qquad  ;\qquad \ket{{\psi _{\textcolor{black}{i}} ^{\textcolor{black}{a}}}} = \textcolor{black}{\hat{a}^\dag} \textcolor{black}{\hat{i}}\ket{\psi _0} \label{eq:psix}
\end{eqnarray} 
with \vspace*{-0.3cm}
\begin{eqnarray}
\vspace*{-0.3cm}z_x = \mathrm{tr}(\bm{\gamma} ^{0x}\bm{\gamma} ^{0x\dag}) = \mathrm{tr}(\bm{\gamma} ^{0x\dag}\bm{\gamma}^{0x}),
\end{eqnarray}
then the 1-TDM elements can be seen as the coefficients corresponding to these singly-excited Slater determinants in the expression of the total excited state
\begin{equation}
 \varphi _i   \; \xrightarrow{\displaystyle{ z_x^{-1/2}(\bm{\gamma}^{0x})_{ia}}} \; \varphi _i \qquad   1 \leq i \leq N < a \leq L
\end{equation}
and one can split the total spinorbital space into two parts (occupied - $i$ indices - and virtual - $a$ indices) and pair the occupied ($\varphi _i$) and virtual ($\varphi _a$) spinorbitals according to the coefficients $z_x^{-1/2}(\bm{\gamma}^{0x})_{ia}$ corresponding to each couple, so the electronic transition can be pictured as a linear combination of hole-to-particle ($1h1p$) contributions, with a total of one electron promoted from the ground to the excited state if we are in the abovementioned conditions. Note that such occupied-to-virtual space mapping through the 1-TDM is not restricted to this type of $\ket{\psi _x}$.
\\ $\;$ \\ \textit{Natural Transition Orbitals} \\ $\;$ \\
Considering the possibility that several occupied/virtual couples can enter the composition of the electronic transition with comparable relative importance but a different topological message, it is possible to change the basis for the representation of the transition into a diagonal one from which one can clearly identify one (sometimes two) hole and one (sometimes two) electron wave functions, so that in most of the cases only one couple of the so-called Natural Transition Orbitals (NTOs) is required for retrieving all the physics of the transition from the excited state calculation.\cite{martin_natural_2003
,batista_et_martin_natural_2004,dreuw_single-reference_2005,mayer_using_2007,surjan_natural_2007,plasser_new_2014,etienne2015transition} In the simplest case (the $\ket{\psi _x}$ defined in equation \ref{eq:psix}), one can shrink the 1-TDM $z_x^{-1/2}\bm{\gamma}^{0x}$ into a rectangular \textbf{T} matrix:
\begin{eqnarray} 
(\bm{\gamma} ^{0x})_{rs} = 0 \; \forall s < N \; \mathrm{and/or} \; r > N \; \Rightarrow \; z_x^{-1/2}\bm{\gamma} ^{0x} \equiv \left( 
 \begin{array}{cc}
0_{o\textcolor{white}{\times o}}  & \textbf{T}_{\textcolor{white}{v}}   \\
0_{v \times o}  &  0_v \\
 \end{array}\right) \quad ; \quad (\textbf{T})_{ic} = z_x^{-1/2}(\bm{\gamma} ^{0x})_{ia},
\end{eqnarray}
with $c = a - N$, $0_o$ and $0_v$ being the zero matrices with the occupied ($N \times N$) or virtual ($[L-N] \times [L-N]$) space dimensions, and $0_{v \times o}$ being the zero matrix with $(L-N) \times N $ dimensions. This operation can be summarized as
\begin{eqnarray}
z_x^{-1/2}(\bm{\gamma}^{0x})\; \in \; \mathbb{R}^{L \times L} \quad \longleftrightarrow \quad \textbf{T} \; \in \; \mathbb{R}^{N \times (L-N)}.
\end{eqnarray}
Following the reduction of the 1-TDM into the rectangular \textbf{T}, one perform its singular value decomposition\cite{amos_single_1961}
\begin{eqnarray} 
\exists \, (\textcolor{black}{\textbf{O}},\!\textcolor{black}{\textbf{V}},{\bm{\lambda}}) \; | \; \textcolor{black}{\textbf{O}^\dag}\textbf{T}\textcolor{black}{\textbf{V}} = \textcolor{black}{\bm{\lambda}}.   \label{svdntos}
\end{eqnarray}
In equation \ref{svdntos}, the \textbf{O} and \textbf{V} matrices contain in their columns the left and right eigenvectors of \textbf{T}. Their components are used for the transformation of the SOs into NTOs:
\begin{eqnarray}
\varphi ^o_i (\textbf{r}_1) = \sum _{j=1}^N (\textbf{O})_{ji} \,\varphi _j (\textbf{r}_1) \qquad \overset{\displaystyle(\bm{\lambda})_{ii}}{\longleftrightarrow} \qquad \varphi ^v_i (\textbf{r}_1) = \sum _{j=1}^{L-N} (\textbf{V})_{ji}\, \varphi _{N+j}(\textbf{r}_1),
\end{eqnarray}
This procedure allows one to condensate the physics of a transition into one couple of one-particle wave functions, with a clear picture of ``where the electron comes from and where it goes".
\\ $\;$ \\ \textit{Natural Difference Orbitals}\\ $\;$ \\
Similar to the NTOs derivation, when diagonalizing the difference density matrix (see equation \ref{eq:MdagM}) instead of the shrunk transition density matrix, it is possible to visualize the so-called Natural Difference Orbitals (NDOs). We can write $\textbf{m}^r(k)$, the $k^{\mathrm{th}}$ component of the $r^{\mathrm{th}}$ natural difference orbital (i.e., the eigenvectors of the difference density matrix\cite{plasser_new_2014,etienne_theoretical_2018}, stored in the \textbf{M} matrix -- see equation \ref{eq:MdagM}) and give the NDOs expression in the real space: 
\begin{equation}
\varphi ^{\Delta}_r (\textbf{r}_1)\;  \overset{\mathbb{R}^3}{\longleftarrow} \; \textbf{m}^r(k) = (\textbf{M})_{kr} \qquad , \qquad  \textbf{M} \; \in \; \mathbb{R}^{L\times L}.
\end{equation}
Note that since the difference density matrix is the square, Hermitian\cite{etienne_theoretical_2018} $- \textbf{TT}^\dag \oplus \textbf{T}^\dag\textbf{T}$ matrix, its diagonalization does not produce left (occupied) and right (virtual) eigenvectors, but a set of $L$ eigenvectors. However, according to the definition of $\ket{\psi _x}$ given in equation \ref{eq:psix} and given the structure of the corresponding 1-TDM, one can deduce\cite{etienne_theoretical_2018} that
\begin{equation}
\bm{\gamma}^d = \textbf{TT}^\dag \oplus 0_v \quad ; \quad \bm{\gamma}^a = 0_o \oplus \textbf{T}^\dag\textbf{T}, \label{eq:gamma_da_oplus}
\end{equation}
i.e.,
\begin{equation}
\textbf{M} = \textbf{O} \oplus \textbf{V} \quad ; \quad \textbf{m} = - \bm{\lambda}\bm{\lambda}^\dag \oplus \bm{\lambda}^\dag\bm{\lambda} \label{eq:Mm_oplus}
\end{equation}
which means that the occupied and virtual NTOs are the eigenvectors of the detachment and attachment 1-RDM, respectively. Note that in equations \ref{eq:gamma_da_oplus} and \ref{eq:Mm_oplus}, the ``$\oplus$" symbol stands for the direct sum of two matrices.
\\ $\;$ \\ \newpage \noindent \textit{The special case of RPA, TDHF, TDDFT and BSE} \\ $\;$ \\
Popular excited states calculation methods such as the Random Phase Approximation (RPA), the Time-Dependent Hartree Fock (TDHF) theory, the Time-Dependent Density Functional Theory (TDDFT) or the GW-Bethe Salpter equation (BSE) being of particular interest for computing molecular electronic transitions, we did extend the strategies for analyzing the electronic transitions topology to these methods. Refs.\cite{PhysRev.115.786,hirata1999configuration,hirata_time-dependent_1999-1,dreuw_single-reference_2005,casida_time-dependent_1996,casida_time-dependent_2009,gui_accuracy_2018} give more details about these methods. They couple the single excitations (with amplitudes stored in an \textbf{x} vector) described above (see equation \ref{eq:psix}) to de-excitations through a \textbf{y} vector, as expressed for example in the equation of motion linear response (LR) transition operator\cite{sauer_molecular_2011}
\begin{equation}
\hat{\mathcal{T}} = \sum _{i=1}^N \sum _{a=N+1}^L \textbf{x}_{ia} \hat{a}^\dag \hat{i} - \textbf{y}_{ia} \hat{i}^\dag \hat{a} 
\label{eq:Top}
\end{equation}
where \textbf{x} and \textbf{y} are the vectors of LR excitation/de-excitation amplitudes. The ``$\mathcal{T}$" symbol in this section will refer to mathematical objects derived from RPA/TDHF/TDDFT/BSE methods. 

Similar to the singular value decomposition of the shrunk transition density matrix described above, it was proposed for this class of methods to build two rectangular matrices and to produce their left and right eigenvectors. The first one consists in a recast of elementary transition tensors norms stored into a $\bm{\tau}$ vector into a rectangular $\overline{\textbf{T}}$ matrix\cite{etienne2015transition}
\begin{equation}
\sqrt{\textbf{x}_{ia}^2 + \textbf{y}_{ia}^2} = \bm{\tau}_{ia} \Rightarrow \bm{\tau} \; \overset{\propto}{\longleftrightarrow} \; \overline{\textbf{T}} \; \in \; \mathbb{R}^{N \times (L-N)} \qquad \Rightarrow \qquad \exists \, (\overline{\textbf{O}}, \overline{\textbf{V}}, \overline{\bm{\lambda}}) \; | \;  \overline{\bm{\lambda}} = \overline{\textbf{O}}^\dag \, \overline{\textbf{T}} \, \overline{\textbf{V}}.
\end{equation}
Here, the columns of $\overline{\textbf{O}}$ and $\overline{\textbf{V}}$ contain the components of occupied $\left\lbrace \overline{\varphi} ^o\right\rbrace$ and virtual $\left\lbrace \overline{\varphi} ^v\right\rbrace$ transition orbitals in the canonical spinorbital space. 

Another attempt to provide a transposition of the NTOs strategy to the RPA/TDHF/TDDFT/BSE methods consisted in considering as the central quantity of interest the relative weight of an occupied/virtual spinorbital couple.\cite{etienne2015transition} From transition operator in equation \ref{eq:Top}, it appears, due to the non-standard normalization condition corresponding to these methods, i.e.,
\begin{equation}
\sum _{i=1}^N \sum _{a=N+1}^L \textbf{x}_{ia}^2 - \textbf{y}_{ia}^2 = 1,
\end{equation}
that the relative weight of an occupied/virtual canonical couple $(\overline{\bm{\vartheta}}_\mathcal{T})_{ia} = \textbf{x}_{ia}^2 - \textbf{y}_{ia}^2$ can lead to the construction of an alternative transition matrix $\textbf{T}_\mathcal{T}$ and to its singular value decomposition, producing the so-called canonical transition orbitals (CTOs)\cite{etienne2015transition}
\begin{equation}
(\overline{\bm{\vartheta}}_\mathcal{T})_{ia} = \textbf{x}_{ia}^2 - \textbf{y}_{ia}^2\; {\longleftrightarrow} \; \textbf{T}_\mathcal{T} \propto (\bm{\vartheta}_\mathcal{T})_{ia}^{1/2} \; \longrightarrow \; \left(\textbf{O}_\mathcal{T}, \textbf{V}_\mathcal{T},  \bm{\lambda}_\mathcal{T} \right) \; \longrightarrow \; \left\lbrace \varphi _\mathcal{T}^o,\varphi _\mathcal{T}^v \right\rbrace
\end{equation}
Note that in practise, in most of the cases the squared contributions of the de-excitations (see below) are relatively small when compared to the excitations in the operator in equation \ref{eq:Top}, so that generally the CTOs and NTOs have very close shapes. For diagnosis purpose, it is possible to assess the relative importance of the hole-particle/particle-hole correlation in such cases\cite{etienne2015transition}
\begin{equation}
\mu ^{0x} = \mathrm{tr}\left(\textbf{YY}^\dag\right) = \dfrac{1}{2}\left( \, \overline{\lambda} - 1\right) \qquad \overline{\lambda} = \mathrm{tr}\left(\bm{\overline{\lambda}\, \overline{\lambda}}^\dag \right).
\end{equation} 
\subsection{Quantitative analysis}
In addition to providing a clear picture of the transition with visual tools based on charge densities or wave functions, topological analyses of molecular electronic transitions can also be quantitative. Many strategies exist for putting a number on the nature of a transition. Among them, some are based on the segmentation of the geometrical structure of the molecule into molecular fragments, and the evaluation of the inter- or intra-fragment local charge density changes. On the other hand, one can think of systematic (i.e., related to the system) approaches providing a quantitative assessment of the locality of a charge transfer, or think in terms of amount of displaced charge for example. As we will also see, from some fragment-based approaches one can derive charge transfer numbers which describe the photoinduced electronic structure reorganization or the molecule in its globality. Additionally, some statistical analysis of the exciton size are also possible, as introduced later in this paragraph. Finally, some quantities, indirectly related to the nature of the charge transfer, will be identified.

Prior to the definition of the topological metrics, let's note that for practical reasons the local basis $\left\lbrace \varphi \right\rbrace$ is very often expanded into another basis (the basis of atomic functions). This is called the Linear Combination of Atomic Orbitals (LCAO) approximation, and it expresses any spinorbital as
\begin{equation}
\varphi _s (\textbf{r}_1) = \sum _{\mu = 1}^K (\textbf{C})_{\mu s}\phi _\mu (\textbf{r}_1) \qquad 1 \leq s \leq L.
\end{equation}
In most of the cases, this basis is not orthogonal, and the spatial overlap between two basis functions is computed and stored into the LCAO overlap matrix elements:
\begin{equation}
(\textbf{S})_{\mu \nu} = \int _{\mathbb{R}^3} d\textbf{r}_1\; \phi _{\mu}^*(\textbf{r}_1)\phi _\nu  (\textbf{r}_1).
\end{equation}
Since in most of the cases the atomic orbitals are centered on the atomic positions, for a given quantum state $\ket{\psi _m}$, there exist several schemes for trying to evaluate the electronic population of these atomic orbitals, i.e., the amount of charge to be attributed to the different atoms, which gives an approximate mapping of the charge distribution in the molecule:
\begin{equation}
q_A^m = \sum _{\mu \in A} \left(\textbf{S}^w \textbf{C}\bm{\gamma}_m\textbf{C}^\dag\textbf{S}^z\right)_{\mu\mu} = \sum _{\mu \in A} \left(\textbf{S}^w \textbf{P}_m\textbf{S}^z\right)_{\mu\mu} \qquad 0 \leq w = 1 - z \leq 1.
\end{equation}
Note that any combination of $w$ and $z$ follows the constraint that the sum over all the fragments of $q_A^m$ will return the total number of electrons of the system, which makes this type of population analysis quite arbitrary. In these conditions, one can then decide to evaluate the amplitude of the population change for any specific fragment when the molecule goes from its ground to an electronic excited state by computing
\begin{equation}
\Delta q_A^{0x} = q_A^x - q_A^0.  
\end{equation}
Aside from this quantity, one can also try to use the exciton wave function for computing a number relating the nature of an electronic transition: Inserting the LCAO expression of the spinorbitals in the projection of the exciton wave function into the local basis $\left\lbrace \varphi \right\rbrace$ gives
\begin{align*}
\tilde{\gamma}^{0x}(\textbf{r}_1;\textbf{r}_1') &= \sum _{r,s=1}^L \varphi _r(\textbf{r}_1) (\bm{\gamma}^{0x})_{rs} \varphi _s^*(\textbf{r}_1') \\
&= \sum _{r,s=1}^L \sum _{\mu ,\nu=1}^K (\textbf{C})_{\mu r}(\bm{\gamma}^{0x})_{rs}(\textbf{C}^{\dag})_{s\nu} \phi _\mu(\textbf{r}_1) \; \phi _\nu ^* (\textbf{r}_1') \\
&= \sum _{\mu ,\nu=1}^K  (\textbf{D}^{0x})_{\mu \nu} \; \phi _\mu(\textbf{r}_1)\phi _\nu^* (\textbf{r}_1')
\end{align*}
where $\textbf{D}^{0x}$ is the 1-TDM in the basis of atomic functions. From this expression of $\tilde{\gamma}^{0x}$ and the definition of \textbf{S} one can evaluate an excitation number as the integral of the product of the exciton wave function with its complex conjugate\cite{plasser_analysis_2012}
\begin{equation}
\Omega^{0x} = \int _{\mathbb{R}^3} d\textbf{r}_1 \; \int _{\mathbb{R}^3} d\textbf{r}_1' \; \tilde{\bm{\gamma}}^{0x}(\textbf{r}_1;\textbf{r}_1')\tilde{\bm{\gamma}}^{0x*}(\textbf{r}_1;\textbf{r}_1')
\end{equation}
that is, in the LCAO basis:
\begin{equation}
\Omega^{0x} = \int _{\mathbb{R}^3} d\textbf{r}_1 \; \int _{\mathbb{R}^3} d\textbf{r}_1' \; \sum _{\mu , \nu =1}^K  \sum _{\sigma , \lambda =1}^K   \;(\textbf{D}^{0x})_{\mu \nu} \; \phi _\mu(\textbf{r}_1)\phi _\nu^* (\textbf{r}_1')(\textbf{D}^{0x})_{\mu \nu}^* \; \phi _\sigma ^*(\textbf{r}_1)\phi _\lambda (\textbf{r}_1')
\end{equation}
and reduces to
\begin{equation}
\Omega^{0x} = \sum _{\mu \nu =1}^K  \sum _{\sigma , \lambda =1}^K  \; (\textbf{D}^{0x})_{\mu \nu} (\textbf{S})_{\mu \sigma} (\textbf{D}^{0x\dag})_{\sigma \lambda} = \mathrm{tr}\left(\textbf{D}^{0x}\textbf{S}\textbf{D}^{0x\dag}\textbf{S}\right).
\end{equation}
This quantity is the number of electrons promoted during the electronic transition process. Obviously, this excitation number is method-dependent, so for example an electronic transition leading to an excited state defined as in equation \ref{eq:psix} will return an excitation number of one, because one electron is promoted in total. On the other hand, transitions generated by the operator in equation \ref{eq:Top} can lead to promotion numbers superior to one. 

Note that if the atomic basis set was orthonormal, the $\Omega^{0x}$ quantity would become\cite{luzanov_interpretation_1980}
\begin{equation}
(\textbf{S})_{\mu \nu} = \delta _{\mu \nu} \; \Rightarrow \; \Omega^{0x} _\perp = \sum _{\mu , \nu=1}^K  \sum _{\sigma , \lambda=1}^K  \; (\textbf{D}^{0x}_\perp)_{\mu \nu}\delta _{\mu \sigma}(\textbf{D}^{0x\dag}_\perp)_{\lambda \sigma}\delta _{\nu \lambda} = \underbrace{\sum _{\mu , \nu=1}^K (\textbf{D}^{0x}_\perp)_{\mu \nu}(\textbf{D}^{0x\dag}_\perp)_{\nu \mu}}_{\displaystyle{\mathrm{tr(\textbf{D}^{0x}_\perp\textbf{D}^{0x\dag}_\perp)}}} = \sum _{\mu , \nu=1}^K \left| (\textbf{D}^{0x}_\perp)_{\mu \nu} \right|^2 \label{eq:preluzanov}
\end{equation}
If we are now interested in isolating one $A \rightarrow B$ inter-fragment contribution to the global charge transfer, we can alter the $\Omega^{0x}$ by restricting the integration over \textbf{r}$_1$ and $\textbf{r}_1'$ locally to atoms $A$ and $B$ respectively, which implies that the sum over $\mu$ and $\sigma$ ($\nu$ and $\lambda$) only holds for indices related to atomic orbitals centered on $A$ ($B$), and that if we want to keep the expression of a fragment-based $\Omega^{0x}$-like index with matrix products, the matrices entering the definition of $\Omega^{0x}$ should be shrunk to cover the atomic spaces span by atoms $A$ and $B$. To this end, a fragment-based index, $\Omega_{AB}^{0x}$, has been introduced\cite{plasser_analysis_2012}
\begin{equation}
\Omega_{AB}^{0x} = \dfrac{1}{2}\sum _{\mu , \sigma \in A}   \sum _{\nu , \lambda \in B}  (\textbf{D}^{0x}_{AB})_{\mu \nu} (\textbf{S}_B)_{\nu \lambda}(\textbf{S}_A)_{\mu \sigma}(\textbf{D}^{0x}_{AB})_{\sigma \lambda} = \dfrac{1}{2}\sum _{\mu \in A} \sum _{\lambda \in B} (\textbf{D}^{0x}_{AB}\textbf{S}_B)_{\mu \lambda}(\textbf{S}_A\textbf{D}^{0x}_{AB})_{\mu\lambda}
\end{equation}
with
\begin{equation}
\textbf{D}^{0x}_{AB} \; \in \; \mathbb{R}^{K_A \times K_B} \qquad ; \qquad \textbf{S}_W \; \in \; \mathbb{R}^{K_W \times K_W} \qquad W = A,B
\end{equation}
so that one can evaluate a global, normalized charge transfer number,\cite{plasser_analysis_2012} $\omega _{\mathrm{CT}}^{0x}$ by accounting for every inter-fragment contributions to the total molecular charge transfer (i.e., the cumulated weight of all the singly-excited Slater determinants for which the hole and the particle are located on different fragments):
\begin{equation}
\omega _{\mathrm{CT}}^{0x} = \dfrac{1}{\Omega^{0x}}\sum _{B \neq A}\sum _{A} \Omega^{0x} _{AB}.
\end{equation}
Similar to the relation between $\Omega^{0x}$ and $\Omega^{0x}_\perp$, one can write an inter-fragment charge transfer number for an orthogonal atomic basis
\begin{equation}
l_{A\rightarrow B}^{0x} = \sum _{\mu \in B} \sum _{\nu \in A} \left|(\textbf{D}^{0x}_\perp)_{\mu \nu} \right|^2
\end{equation}
which was proposed by Luzanov few decades ago.\cite{luzanov_interpretation_1980} Note that this $A \rightarrow B$ contribution corresponds schematically to a $\psi _{A^+\!B^-}$ situation, where the charge density is displaced from fragment $A$ to fragment $B$. Following a similar procedure, one can also compute $l_{B \rightarrow A}$ and $l_{A\rightarrow A} = l_A$ indices corresponding respectively to the $\psi _{A^-\!B^+}$ and to the $\psi _{A^*\!B}$ (local excitation on $A$) situations respectively, and derive two quantum metrics,\cite{luzanov_interpretation_1980}
\begin{equation}
L_A^{0x} = l_A^{0x} + \dfrac{1}{2}\left(l_{A\rightarrow B}^{0x} + l_{B \rightarrow A}^{0x} \right) \qquad ; \qquad \Delta D_A^{0x} = \sum _{B \neq A} \left( l_{B \rightarrow A}^{0x} - l_{A \rightarrow B}^{0x}\right),
\end{equation}
the first one being an assessment of the so-called total fragment excitation localization (the summed contribution of local excitation and incoming/outcoming charge density fluctuations), while the second one is a measure of the net change of electron density on fragment $A$, as Luzanov did interpret the two components of $\Delta D_A^{0x}$ as ``charges of opposite sign flowing between $A$ and $B$ on excitation". Here, the sign convention we used leads to the interpretation of $\Delta D_A^{0x}$ as the quantity related to the charge \textit{gained} by fragment $A$ upon the electronic transition.

In addition to the computation of $\omega _{\mathrm{CT}}^{0x}$, it is also possible to evaluate participation ratios\cite{plasser_analysis_2012} in order to count the number of fragments entering a given transition. For this purpose, it is first necessary to evaluate the ratio of fragments entering into the composition of the hole ($\mathrm{PR}_{\mathrm{I}}^{0x}$) and the particle ($\mathrm{PR}_{\mathrm{F}}^{0x}$) before averaging them into $\mathrm{PR}^{0x}$
\begin{equation}
\mathrm{PR}_{\mathrm{I}}^{0x} = \dfrac{(\Omega ^{0x})^2}{\sum _A \left(\sum _B \Omega _{AB}^{0x}\right)^2} \qquad ; \qquad \mathrm{PR}_{\mathrm{F}}^{0x} = \dfrac{(\Omega ^{0x})^2}{\sum _B \left(\sum _A \Omega _{AB}^{0x}\right)^2} \qquad ; \qquad \mathrm{PR}^{0x} = \dfrac{1}{2}\left( \mathrm{PR}_{\mathrm{I}}^{0x} + \mathrm{PR}_{\mathrm{F}}^{0x}\right)
\end{equation}
Note that in the definition of $\mathrm{PR}_{\mathrm{I}}^{0x}$ and $\mathrm{PR}_{\mathrm{F}}^{0x}$ it was arbitrarily chosen that fragment $A$ represents the fragment on which all the initial (``I") orbitals lie, while $B$ is the fragment on which all the final (``F") orbitals lie. While the $\mathrm{PR}^{0x}$ quantity does not inform about the balance of locally excited nature and charge transfer nature for a given excitation, the COH descriptor, expressed as
\begin{equation}
\mathrm{COH} = \dfrac{1}{\mathrm{PR}} \dfrac{\left( \Omega ^{0x}\right) ^2}{\sum _A \sum _B \left(\Omega ^{0x}_{AB}\right)^2},
\end{equation}
has been designed and can provide this information.\cite{plasser_analysis_2012} Note that one can also compute $\mathrm{PR}^{0x}$ by combining $\Omega ^{0x}$ to the 1-TDM singular values\cite{plasser_analysis_2012} contained in $\bm{\lambda}$:
\begin{equation}
\mathrm{PR}_{\mathrm{NTO}}^{0x} = \dfrac{\left(\Omega ^{0x}\right)^2}{\sum _i \left(\bm{\lambda}\right)_{ii}^4}.
\end{equation}
This expression can even be simplified by writing $\mathrm{PR}_{\mathrm{NTO}}^{0x}$ solely as a combination of $\bm{\lambda}$ elements: If we express $\Omega ^{0x}$ in the canonical space,
\begin{equation}
\Omega ^{0x} = \int _{\mathbb{R}^3} d\textbf{r}_1 \; \int _{\mathbb{R}^3} d\textbf{r}_1'\; \sum _{r , s=1}^L \varphi _r (\textbf{r}_1) (\bm{\gamma}^{0x})_{rs} \varphi _s ^* (\textbf{r}_1') \sum _{p , q=1}^L  \varphi _p ^* (\textbf{r}_1) (\bm{\gamma}^{0x})^*_{pq} \varphi _q (\textbf{r}_1') 
\end{equation}
which is
\begin{equation}
\Omega ^{0x} =  \sum _{r , s=1}^L  \sum _{p , q=1}^L   \underbrace{\int _{\mathbb{R}^3} d\textbf{r}_1 \; \varphi _r (\textbf{r}_1) \varphi _p ^* (\textbf{r}_1)}_{\delta_{rp}} \underbrace{\int _{\mathbb{R}^3} d\textbf{r}_1'  \varphi _s ^* (\textbf{r}_1') \varphi _q (\textbf{r}_1')}_{\delta sq}   (\bm{\gamma}^{0x})_{rs}  (\bm{\gamma}^{0x})^*_{pq} = \sum _{r , s=1}^L  (\bm{\gamma}^{0x})_{rs}(\bm{\gamma}^{0x})^*_{rs}.
\end{equation}
Due to the structure of $\bm{\gamma}^{0x}$, it becomes
\begin{equation}
\Omega ^{0x} = \mathrm{tr}(\bm{\gamma}^{0x}\bm{\gamma}^{0x\dag}) = \mathrm{tr}(\textbf{TT}^\dag).
\end{equation}
According to equation \ref{svdntos}, we can rewrite the \textbf{T} and its adjoint as the reverse singular value decomposition:
\begin{equation}
\textbf{T} = \textbf{O}\bm{\lambda}\textbf{V}^\dag \qquad ; \qquad \textbf{T}^\dag = \textbf{V}\bm{\lambda}^\dag\textbf{O}^\dag.
\end{equation}
Therefore, the product of \textbf{T} to the right by its adjoint is a square matrix whose eigenvectors are the left eigenvectors entering the singular value decomposition of \textbf{T}
\begin{equation}
\textbf{TT}^\dag = \textbf{O}\bm{\lambda}\textbf{V}^\dag\textbf{V}\bm{\lambda}^\dag\textbf{O}^\dag \; \Leftrightarrow \; \textbf{O}^\dag \textbf{TT}^\dag \textbf{O} = \bm{\lambda \lambda}^\dag.
\end{equation}
Since the trace of a matrix is an unitary invariant, we find that
\begin{equation}
\mathrm{tr}(\textbf{TT}^\dag) = \mathrm{tr}(\bm{\lambda \lambda}^\dag) \; \Rightarrow \; \Omega ^{0x} = \mathrm{tr}(\bm{\lambda \lambda}^\dag)
\end{equation}
so that $\mathrm{PR}_{\mathrm{NTO}}^{0x}$ becomes
\begin{equation}
\mathrm{PR}_{\mathrm{NTO}}^{0x} = \dfrac{\left(\sum _i \left(\bm{\lambda}\bm{\lambda} ^\dag \right)_{ii}\right)^2}{\sum _i \left(\bm{\lambda}\bm{\lambda} ^\dag\right)_{ii}^2} 
\end{equation}
which is equal to Luzanov's collectivity measure metric,\cite{luzanov_electron_2010} $\kappa ^{\mathrm{CIS}}$.

Another interesting characterization of the electronic transition is given by the computation of global charge transfer numbers related to the nature of the transition.\cite{le_bahers_qualitative_2011} Starting from the state densities, one can for example assess the amount of charge transferred during a transition

\begin{eqnarray}
q_{\mathrm{CT}} = \dfrac{1}{2}\sum _{s = +,-} \int _{\mathbb{R}^3} d\textbf{r} \; n_{s}(\textbf{r}) 
\end{eqnarray}
which should not be confused with the amount of charge displaced during the transition. This distinction is related to the difference between the hole and particle (i.e., detachment/attachment) densities and the difference densities mentioned above.\cite{etienne_new_2014}

In addition to $q_{\mathrm{CT}}$, one can also compute the distance between the centroid of $n_-$ and $n_+$. For this purpose, we define a position vector\cite{le_bahers_qualitative_2011} $\bm{\xi} = (\xi _1 , \xi _2, \xi _3) = (x,y,z)$ and we evaluate its expectation value by integrating the product of its components with $n_-$ and $n_+$
\begin{equation}
\xi _{i,s}^{0x} = q_{\mathrm{CT}} ^{-1} \int _{\mathbb{R}^3} d\textbf{r}_1 \, n_s (\textbf{r}_1) \xi _i \qquad i=1,2,3 \quad ; \quad s = \pm. \label{eq:xipm}
\end{equation}
The three components of the two centroid position vectors are then compiled into the $\textbf{R}_\pm$ vectors, and the norm of the difference between these two vectors represents the distance between the two centroids in the Euclidean space\cite{le_bahers_qualitative_2011}:
\begin{equation}
\textbf{R}_\pm = (\xi _{1,\pm}^{0x}, \xi _{2,\pm}^{0x}, \xi _{3,\pm}^{0x}) \; \Rightarrow \; D_{\mathrm{CT}} = \left|\textbf{R}_+ - \textbf{R}_- \right| = \left\lbrace \sum _{i=1}^3 \left(\xi _{i,+}^{0x} - \xi _{i,-}^{0x}\right)^2\right\rbrace^{1/2} \label{eq:Rpm}
\end{equation}
Note also that the product of the charge transferred by the charge transfer distance gives the norm of a ``charge transfer dipole moment"\cite{le_bahers_qualitative_2011}:
\begin{equation}
\left|\bm{\mu} _{\mathrm{CT}} \right| = q_{\mathrm{CT}}D_{\mathrm{CT}}.
\end{equation}
Besides the charge transfer numbers derived from state densities, one can also use the detachment/attachment densities for quantitatively probing the nature of an electronic transition. For instance, the locality of the charge displacement can be evaluated through the computation of the hole/particle spatial overlap\cite{etienne_toward_2014}
\begin{equation}
\textcolor{black}{\phi _S} = \vartheta _x ^{-1} \, \int _{\mathbb{R}^3} \!\! d\textbf{r}\, \sqrt{\textcolor{black}{n _d (\textbf{r})} \textcolor{black}{n _{a}(\textbf{r})}} \; \in \; [0;\!1] \qquad ; \qquad \vartheta _x = \dfrac{1}{2}  \sum _{q = \textcolor{black}{d},\textcolor{black}{a}}\int _{\mathbb{R}^3} \!\!d \textbf{r} \, n _{q}(\textbf{r}) 
\end{equation} 
This quantity, combined with the normalized charge transferred\cite{etienne_probing_2015}
\begin{eqnarray}
\dfrac{\vartheta _x ^{-1}}{2}\sum _{s = +,-} \int _{\mathbb{R}^3} d\textbf{r} \; n_{s}(\textbf{r}) = \tilde{\varphi} \; \in \; [0;\!1]
\end{eqnarray}
gives a global quantum metric jointly relating the locality of the exciton and the amount of charge transferred during the transition\cite{etienne_probing_2015}
\begin{eqnarray}
\psi = 2 \pi^{-1}\,\underbrace{\mathrm{arctan}\left(\dfrac{\phi _S}{\tilde{\varphi}}\right)}_{\theta _S} = \dfrac{2\theta _S}{\pi} \; \in \; [0;\!1[ \label{eq:psi}.
\end{eqnarray}
where $\theta _S$ is the angle between the real axis of a complex plane and the projection of $\phi _S$ and $\tilde{\varphi}$ in that plane \textcolor{black}{(see Ref.\cite{etienne_probing_2015} for more details)}. Note that, as in the derivation of the intercentroid distance between the net depleted/accumulated charge densities (see equations \ref{eq:xipm} and \ref{eq:Rpm}), one can also compute the distance between the centroid of detachment and attachment densities. This quantity, written $\zeta$ in \textcolor{black}{Ref.\cite{etienne_new_2014}, deviates from $D_{\mathrm{CT}}$ when the overlap between the hole and the particle increases (see figure 2 from Ref.\cite{etienne_new_2014}).}

It is worth pointing out that, since the difference between the attachment and detachment density matrices returns the difference density matrix, and due to the unequivocality of the density matrix-to-charge density mapping, any function or quantity involving $n_+$ or $n_-$ (the $q_{\mathrm{CT}}$ and $D_{\mathrm{CT}}$ numbers for example) could be equivalently derived from state densities or from detachment/attachment 1-CD.\cite{etienne_new_2014} However, it has been demonstrated that the integral of the detachment (or attachment) 1-CD is an upper bound to the one of $n_\pm$, which justifies its use as a normalization term for $\tilde{\varphi}$ (where the ``$\sim$" symbol above $\varphi$ in $\tilde{\varphi}$ was used to notify that the $n_\pm$ functions used for computing it were derived from detachment/attachment density difference, though this detail was dispensable).\cite{etienne_new_2014,etienne_probing_2015} 

Note also that in parallel to the hole/particle spatial overlap ($\phi _S$), one can also compute the hole/particle entanglement entropy\cite{plasser2016entanglement} $S_{\mathrm{H}|\mathrm{E}}$ which is directly related to the effective number of entangled states\cite{plasser2016entanglement} $Z_{\mathrm{HE}}$
\begin{equation}
S_{\mathrm{H}|\mathrm{E}} = - \sum _i (\bm{\lambda}\bm{\lambda}^\dag) _{ii} \, \mathrm{log}_2 (\bm{\lambda \lambda}^\dag)_{ii} \qquad ; \qquad Z_{\mathrm{HE}} = \left( \prod _{i} (\bm{\lambda \lambda}^\dag)_{ii}^{(\bm{\lambda \lambda} ^\dag)_{ii}}\right) ^{-1} = 2^{S_{\mathrm{H}|\mathrm{E}}}
\end{equation}
Besides the derivation of charge transfer numbers from the manipulation and integration of 1-CD functions, one can also use one-particle wave functions to gain quantitative insights into the nature of electronic transitions. For example, a $\Lambda$ descriptor was designed by Tozer and co-workers in order to perform a diagnostic test of exchange-correlation functionals for TDDFT calculations. This descriptor writes\cite{peach_excitation_2008}
\begin{equation}
\Lambda = \dfrac{\displaystyle{\sum _{i=1}^N \sum _{a=N+1}^L \kappa _{ia}^2 O_{ia}}}{\displaystyle{\sum _{i=1}^N \sum _{a=N+1}^L \kappa _{ia}^2 }} \qquad \kappa _{ia} = \textbf{x}_{ia} + \textbf{y}_{ia} \qquad ; \qquad O_{ia} = \int _{\mathbb{R}^3} d\textbf{r}_1 \; \left| \varphi _i (\textbf{r}_1) \right| \left| \varphi _a (\textbf{r}_1) \right|
\end{equation}
and was related to the error observed on the transition energies of molecules, when computed with different types of exchange-correlation functionals. The $\kappa$ elements were also used for constructing another charge transfer metric, $\Delta r$
\begin{equation}
\Delta r = \dfrac{\displaystyle{\sum _{i=1}^N \sum _{a=N+1}^L \kappa _{ia}^2 \left| \braket{\varphi _a | \textbf{r}_1 | \varphi _a} - \braket{\varphi _i | \textbf{r}_1 | \varphi _i} \right| }}{\displaystyle{\sum _{i=1}^N \sum _{a=N+1}^L \kappa _{ia}^2 }}
\end{equation}
related to the spatial structure of the exciton (more particularly, the space covered during the excitation).\cite{GuidoJChemTheoryComput2013} As when we wrote $\mathrm{PR}^{0x}$ using the singular values of the 1-TDM, the $\Delta r$ metric was also transposed to the NTO basis where $\kappa$ is replaced by $\bm{\lambda}\bm{\lambda} ^\dag$, and the occupied and virtual canonical orbitals were replaced by the occupied and virtual NTOs for constructing the $\Delta r_{\mathrm{NTO}}$ index.\cite{guido_effective_2014}

Independently from the development of $\Delta r$, $D_{\mathrm{CT}}$ and $\zeta$, two descriptors were also designed\cite{plasser2015statistical} for relating the root-mean-square deviation separation between the hole and the particle ($d_{\mathrm{exc}}$) and relative electron/hole separation distance ($d_{h \rightarrow e}$):
\begin{equation}
d_{\mathrm{exc}} = \sqrt{\left\langle \left| \textbf{r}_1 - \textbf{r}_1' \right|^2\right\rangle } \qquad ; \qquad
d_{h \rightarrow e} = \left| \braket{\textbf{r}_1' - \textbf{r}_1} \right| 
\end{equation}
with $d_{h \rightarrow e} \leq d_{\mathrm{exc}}$. Note that when evaluating these two expectation values, the normalized integration is performed over the hole and the particle coordinates. Such integral for any operator $\hat{\mathcal{O}}$ gives\cite{bappler_exciton_2014}
\begin{equation}
\braket{\hat{\mathcal{O}}} = \dfrac{1}{\Omega ^{0x}} \int _{\mathbb{R}^3} d\textbf{r}_1 \; \int _{\mathbb{R}^3}d\textbf{r}_1' \; \tilde{\gamma}^{0x}(\textbf{r}_1,\textbf{r}_1')  \, \hat{\mathcal{O}} \,\tilde{\gamma}^{0x*}(\textbf{r}_1,\textbf{r}_1').
\end{equation}
Furthermore, it is also possible to estimate the covariance between the hole and the particle position vectors\cite{plasser2015statistical}
\begin{equation}
\mathrm{COV}(\textbf{r}_1,\textbf{r}_1') = \braket{\textbf{r}_1 \cdot \textbf{r}_1'} - \braket{\textbf{r}_1}\cdot\braket{\textbf{r}_1'}
\end{equation} 
as well as the Pearson correlation coefficient applied to the hole/particle statistical spatial distribution\cite{plasser2015statistical}
\begin{equation}
 R_{eh} = \sigma _{\textbf{r}_1}^{-1}\sigma _{\textbf{r}_1'}^{-1}\,\mathrm{COV}(\textbf{r}_1,\textbf{r}_1')
\end{equation}
 which is the covariance normalized with the standard deviations of the hole/particle positions\cite{plasser2015statistical}
\begin{equation}
\sigma _{\textbf{r}_j} = \sqrt{\braket{\textbf{r}_j ^2} - \braket{\textbf{r}_j}^2} \qquad \textbf{r}_j = \textbf{r}_1,\textbf{r}_1'.
\end{equation}
The $\sigma$ function is also employed for completing the above mentioned $\Delta r$ with a statistical descriptor, $\Delta \sigma$, defined as\cite{guido_effective_2014}
\begin{equation}
\Delta \sigma = \dfrac{\displaystyle{\sum _{i=1}^N \sum _{a=N+1}^L \kappa _{ia}^2 \left| \sigma _a - \sigma _i\right|}}{\displaystyle{\sum _{i=1}^N \sum _{a=N+1}^L \kappa _{ia}^2 }} \qquad ; \qquad \sigma _p = \sqrt{\braket{\varphi _p | \textbf{r}_1 ^2 | \varphi _p} - \braket{\varphi _p || \textbf{r}_1|| \varphi _p}^2}
\end{equation}
which is then simply added to $\Delta r$ in order to give\cite{guido_effective_2014}
\begin{equation}
\Gamma = \Delta r + \Delta \sigma.
\end{equation}
Note finally that this metric can also be derived using the natural transition orbitals rather than the canonical orbitals.\cite{guido_effective_2014}

\section{Electronic structure and charge generation properties of dye-sensitized metal oxides}
\subsection{Dye adsorption and energy levels line-up: electrostatic and charge transfer}
The primary dye/semiconductor interactions are mediated by the dye adsorption mode onto the semiconductor surface, which directly 
influences the interfacial electrostatic and the electronic coupling between the molecule and the substrate, yielding sizeable 
changes in their relative energy level alignment. 
A crucial characteristic for efficient dyes is the presence of suitable functional groups able to strongly bind 
to the semiconducting oxide surface. To favor ultrafast electron injection, the anchoring group indeed 
should coincide, or be very close (conjugated), to the dye acceptor unit, where 
the photo-excited electrons are spatially confined. This promotes electronic coupling between the 
donor levels of the excited dye and the delocalized acceptor levels of the semiconductor 
conduction band, and helps the charge injection process.\cite{GUPTA2015,Lasser-2015,Calbo-2014} 
The sensitizer's anchoring group should also provide stable grafting 
of the dye onto semiconductor surface, thus leading to long-term device stability.\\

The anchoring mechanism of the largely employed carboxylic acid group to the 
TiO$_2$ surface can be exemplified referring to the coordination modes of the 
carboxylate fragment (COO$^-$) to metal ions; there are basically three typical 
coordination schemes: monodentate, bidentate chelating and bidentate bridging, with the last one usually
disregarded since it is less stable. 
Along with experimental investigations, a number of theoretical studies on the dye 
adsorption modes on the TiO$_2$ surface have been published,\cite{AnselmiPhysChemChemPhys2012,
Shklover-1998,DeAngelisNanoLett2007,Schiffmann-2010,DeAngelis-2010,Martsinovich-2010,
Persson-2000,VittadiniJPhysChemB2000,Labat-2011,Labat-2012} starting from the pivotal 
work by Vittadini et al. on the formic acid adsorption on the TiO$_2$ anatase (101) surface.\cite{VittadiniJPhysChemB2000} 
In some cases, calculations showed that for organic dyes bearing a carboxylic acid as anchoring group, 
the preferred adsorption mode was bidentate bridging, with one proton 
transferred to a nearby surface oxygen, 
while the monodentate anchoring is usually predicted to be less stable, 
although some dependency of the relative stability of these two anchoring modes on the employed 
computational methodology has been outlined.\cite{MosconiSelloni-2012,VittadiniJPhysChemB2000}

The energetics of the TiO$_2$ conduction band are known to depend on several factors, such
as the local pH,\cite{RothenbergerJPhysChem1992,OReganChemPhysLett1991,
BoschlooJPhysChemB1999} the concentration of potential determining
ions ({\itshape e.g.} Li$^+$)\cite{BoschlooJPhysChemB1999,RedmondJPhysChem1993} 
and, possibly in relation to acid-base equilibria,
also on the nature of the electrolyte 
solvent.\cite{RedmondJPhysChem1993,EnrightJPhysChem1994} 
The role of surface adsorbed molecules, including the dye, in determining
the TiO$_2$ CB energetics is much less clear.\cite{RuhleJPhysChemB2004,
ArdoChemSocRev2009,WestermarkChemPhys2002,RuhleJPhysChemB2005,YanJPhysChem1996,
DeAngelisNanoLett2007,PastorePhysChemChemPhys2012,KusamaLangmuir2008,
TachibanaJPhysChemB2001}
Different works on ruthenium dyes have shown a correlation between the
dye protonation state and the DSC performance,\cite{YanJPhysChem1996,
DeAngelisNanoLett2007,TachibanaJPhysChemB2001} with dyes
carrying a higher number of protons leading to higher photocurrent values, $J_{sc}$, but
lower open circuit voltage, $V_{oc}$. An interesting correlation between the dipole
moment of co-adsorbing species, mainly substituted benzoic
acids, and the corresponding DSC $V_{oc}$ was observed by R\"uhle
{\itshape et al.},\cite{RuhleJPhysChemB2005} 
who pointed out a linear relationship between the dye
coverage ($N$), the dipole ($\mu$) component normal to the surface
($\theta$ is the molecule tilting angle) and the potential shift ($\Delta V$) at
the interface affecting the TiO$_2$ CB energy:
\begin{equation}\label{eqn:DeltaV}
\Delta V=\frac{N\mu\cos\theta}{\varepsilon\varepsilon_0}
\end{equation}
where $\varepsilon$ is the dielectric constant of the dye monalayer and $\varepsilon_0$
is the dielectric permittivity of vacuum.
For ruthenium dyes, a correlation between the
dye adsorption mode and the corresponding DSC $V_{oc}$
has been observed.\cite{DeAngelisNanoLett2007} 
Later works on solid-state DSC
has clearly shown a greater than 100 mV TiO$_2$ conduction band
shift between a heteroleptic ruthenium dye and an organic dye,
which was interpreted in terms of a dipole-induced TiO$_2$ CB
shift of different sign.\cite{ChenNanoLett2009}
Such shifts are generally more difficult
to be observed in DSC based on a liquid electrolyte,\cite{MiyashitaJAmChemSoc2008} 
in which the high ion strength and the effect of thermal motion may
hinder the role of interface dipoles. Nevertheless, Kusama {\itshape et al.}
reported a combined experimental and theoretical study which
showed a clear correlation between the dipole moment of
electrolyte additives and their DSC $V_{oc}$.\cite{KusamaLangmuir2008}

As a matter of fact, when a dye binds to a semiconductor surface, two effects
might be at work: (a) the aforementioned electrostatic (EL)
effect, due to the dye dipole moment; and (b) the effect of the
charge transfer (CT) between the dye and the semiconductor,
which may accompany the dye-semiconductor bond formation.
We have used the so-termed Charge Displacement (CD) analysis,\cite{BelpassiJAmChemSOc2008} 
to investigate the adsorption of several prototypical organic dyes and co-adsorbents on TiO$_2$ models, 
quantifying and rationalizing the effects of EL and CT contributions to the TiO$_2$ CB energetics.
\cite{RoncaEnergyEnvironSci2013} 

Similar to the ESCD strategy (see equation \ref{eqn:deltaq0x}), one can analyze the electron density rearrangement by defining the charge displacement (CD) along the \textit{z} direction as: 

\begin{equation}
\label{eqn:deltaq}
\Delta q(z)=\displaystyle
\int_{-\infty}^{z} dz'
\int_{-\infty}^{+\infty}
\int_{-\infty}^{+\infty}
\Delta \rho(x,y,z')\,dx\,dy. 
\end{equation}

where $\Delta \rho$ is the electron density difference and $\Delta q$ measures, at each point along the $z$ axis, the electron
charge that, upon formation of the adduct, is transferred from the right to the left side of the perpendicular plane through z (a negative value thus corresponds to electron flow from left to right).  

\begin{figure}[!htp]
\centering
\includegraphics[angle=90,width=8cm]{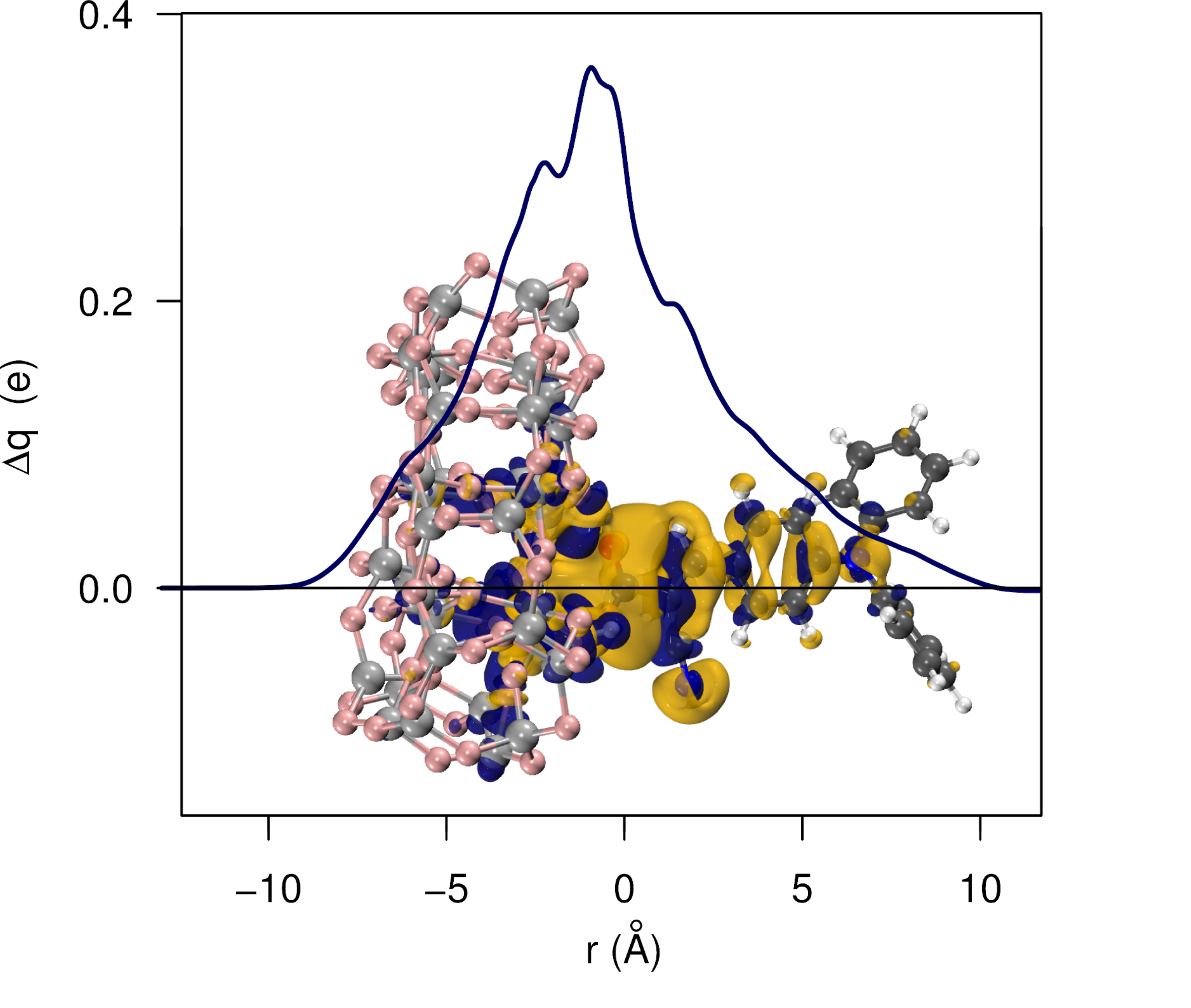}
\caption{\label{fig:CD_L0}{\small Isodensity contour plot and charge displacement curve for L0 adsorbed
onto TiO$_2$ in a BB configuration. Yellow surfaces identify regions in which the
electron density decreases whereas zones of density accumulation are marked by
dark blue surfaces. The density value at the surfaces is $\pm$0.0005 e au$^{-3}$.} Reprinted from Ref.
\cite{RoncaEnergyEnvironSci2013} by permission of The Royal Society of Chemistry.}
\end{figure}

In this case, after defining the average TiO$_2$ surface
plane, we choose as integration direction ($z$) that 
perpendicular to this plane and passing through the 
carboxylic carbon on the anchoring group of the dye. 
The dye-TiO$_2$ geometry was fully relaxed in \emph{vacuo} using
the PBE exchange correlation functional with the ADF program package;\cite{ADF2001} 
a TZVP(DZVP) basis set for Ti (H, C, N, O, S) atoms was adopted. 
The electron density was obtained by a DFT/B3LYP
single point calculation employing the SVP basis set and the solvent (acetonitrile) have been 
described by means of the C-PCM solvation model\cite{cpcm2} as implemented
in Gaussian 09 (G09).\cite{g09}
In Figure \ref{fig:CD_L0} 
we show the isodensity contour plot of the electron 
density difference and the
CD curve, calculated by equation~\ref{eqn:deltaq} for the prototypical L0 dye
adsorbed onto a (TiO$_2$)$_{38}$ cluster in the bidentate bridging (BB) configuration. 
We can see significant charge depletion 
lobes in the areas around the carboxylic group, the nitrogen atom of the nitrile group and 
the phenyl bound to the cyanoacrylic moiety. Charge accumulation lobes can be seen instead on 
the central carbon of the cyanoacrylic anchoring group of the sensitizer, 
in proximity of the proton detached from the dye and adsorbed on the TiO$_2$ surface and, in particular, 
in regions of the semiconductor cluster. Note that the curve is largely positive across the entire adduct, 
implying a continuous charge transfer from the dye to the TiO$_2$. The maximum of the CD curve in the inter-fragment region is 0.36 electrons. 
Different dyes adsorbed in the same bidentate adsorption mode showed little variations in the amount of CT, 
while the molecular monodentate adsorption geometry yielded to a much smaller CT.

Assuming that this marked charge redistribution, along with a pure electrostatic (EL) contribution due to the dye dipole, 
induce significant modifications on the TiO$_2$ CB edge, one can formulate a simple interpretative model 
by expressing the total shift, $\Delta CB_{TOT}$, as the sum of the two main effects strictly related to the dye sensitizer 
($\Delta CB_{TOT}=\Delta CB_{EL} + \Delta CB_{CT}$).

Concerning the EL effect, we evaluated the CB shift in relation to the electrostatic potential generated by 
the dye molecule rather than with its standalone dipole moment. This quantity represents the effective average 
electrostatic potential generated by the dye charge distribution in the region of the first semiconductor titanium layer. 
We demonstrated that the electrostatic potential generated by the dye charge distribution correlates 
linearly and very accurately with the observed TiO$_2$ conduction band shift. 
This implies that a direct correlation between the dye dipole and the observed conduction band 
shift effectively exists only for dyes of similar structure and dimensions. 
The estimated electrostatic contribution to the conduction band shift amounts to about 40\% of the total calculated shifts. 
Considering the pronounced ground state dye/semiconductor charge transfer, we investigated the supposition 
that the remaining contribution to the semiconductor CB shift is directly due to charge transfer effects. 
We established that there is indeed near-exact proportionality between the amount of charge transfer 
calculated by the charge displacement analysis and the residual contribution to the conduction band shift. 
We thus found that the CT induced CB shift may be as large as 60\% of the total shift.\\ 

\begin{figure}[h!]
\centering
\includegraphics[height=40mm, angle=0]{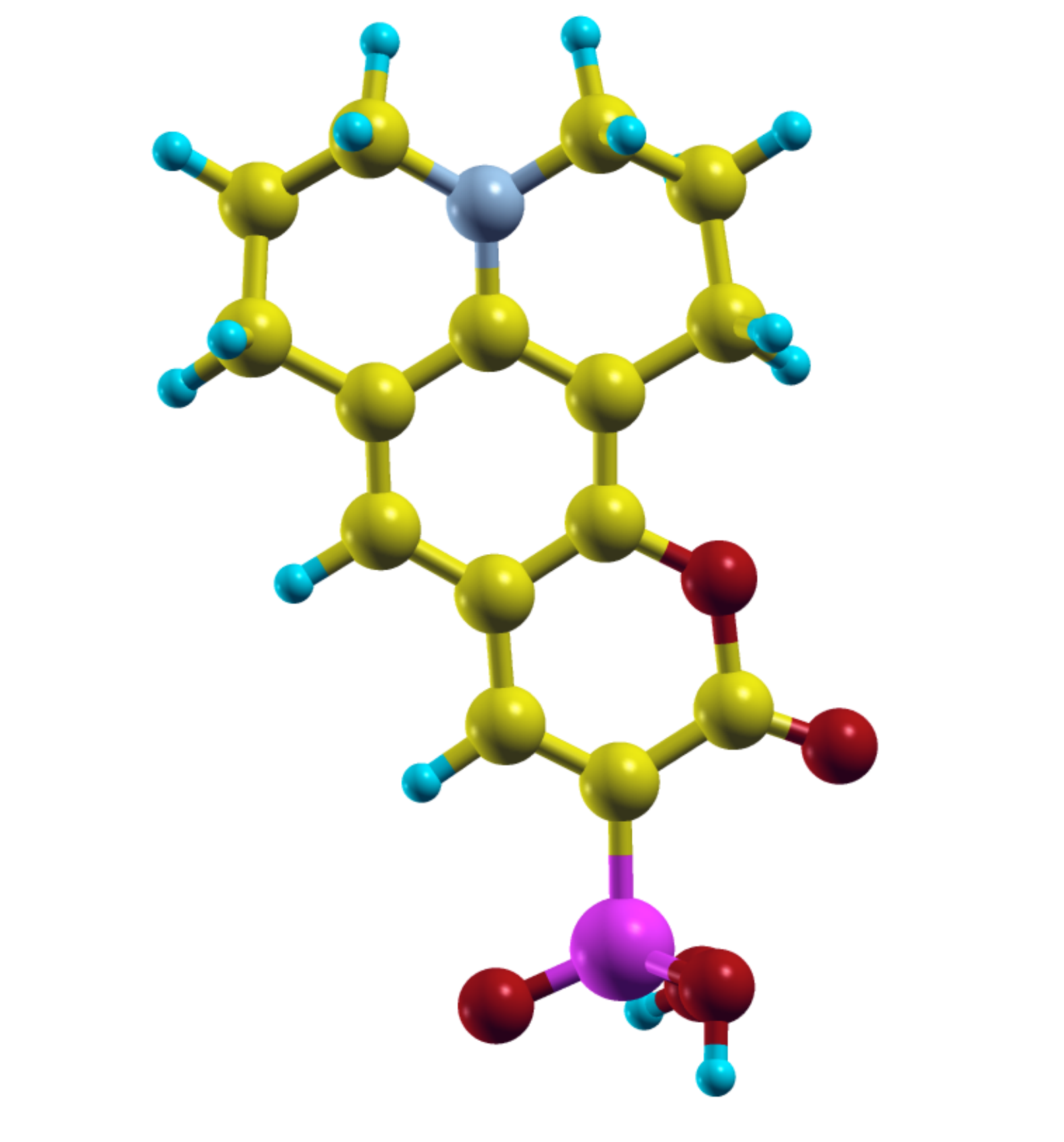}
\includegraphics[height=40mm, angle=0]{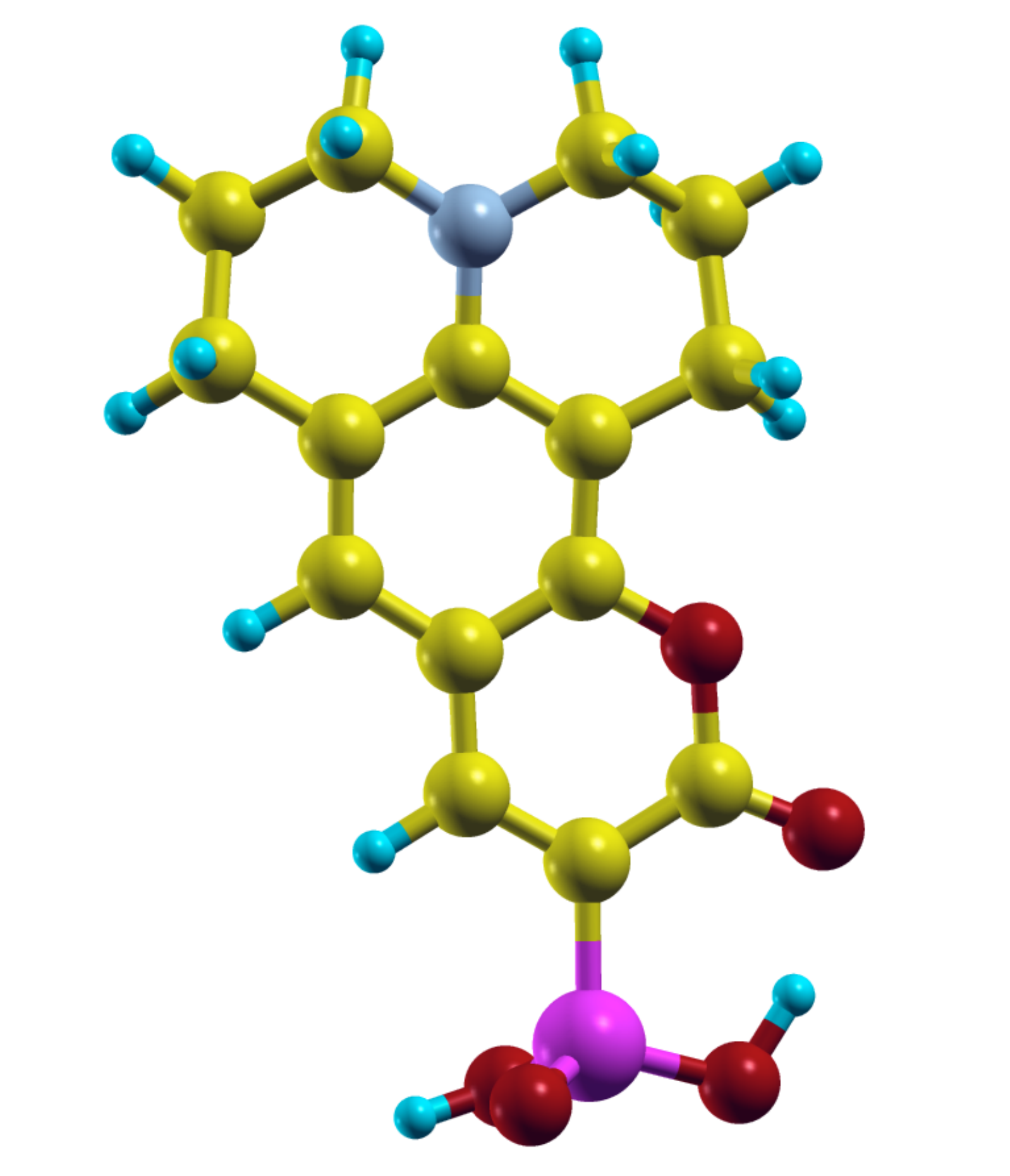}
\caption{\label{Fig_C343} Optimized geometries of two isomers of the 
C343 dye molecule with a phosphonic acid anchoring group. Left: H-down. Right: H-up.}
\end{figure}

While hybridization of the dye/semiconductor molecular orbitals traslates into a considerable net charge transfer from the dye
to the TiO$_2$ CB, the electrostatic effect dominates in the case of decoupled subsystems, as we shall discuss 
below for the cumarin C343-sensitized nickel oxide surface.\cite{Piccinin-2017}   
Contrary to the rich literature on the functionalization of TiO$_2$ substrates by molecular sensitizers,\cite{Persson-2000,DeAngelis-2007,Troisi-2011,Agrawal-2013,LeBahers-2013,Agrawal-2013,Pastore-2014,Mosconi-2012,Monti-2015,Pastore-2017} 
the characterization of the electronic and structural properties of dye-sensitized NiO interfaces in both $p$-type DSCs and water reduction photocathodes  
is still modest. In the few theoretical work reported so far,\cite{Munoz-Garcia2015,Rantala-2016,Wykes-2016, Piccinin-2017} 
the energy level alignment across the interface, and in particular the relative position of the dye's HOMO 
with respect to NiO's VB, ruling the hole injection driving force, was found to be extremely sensitive to the anchoring group  
and binding mode (monodentate, bidentate, tridentate) of the dye. We have recently characterized the interaction of 
a derivative the cumarin C343 dye, featuring a terminal phosphonic acid (-PO(OH)$_2$) anchoring group and the (100) NiO surface. 
We have considered two possible conformers:\cite{Weppeng-2012,Cave-2002} one where one of the H atoms of the phosphonate is H-bonded 
to the oxygen of the carbonyl group of the coumarin (H-up), the other where both H atoms of the phosphonate point away from it (H-down) (see Figure \ref{Fig_C343}). The DFT calculations (Quantum ESPRESSO software \cite{QEpaper})
were carried out using the PBE exchange and correlation functional, 
with a Hubbard U term of 4.6 eV. \cite{Cococcioni2005}. 
We used ultrasoft pseudopotentials \cite{USPP} and a plane waves basis set, with a cutoff of 
30 Ry for the wave functions and 300 Ry for the charge density. Dispersion corrections to DFT were 
modeled using Grimme's method \cite{JCC:JCC20495}. Further details can be found in the original paper\cite{Piccinin-2017}.
In gas phase, the H-bonded isomer is slightly favored, by 0.20 eV using the PBE functional.

\begin{figure}[ht!]
\centering
\includegraphics[width=100mm, angle=0]{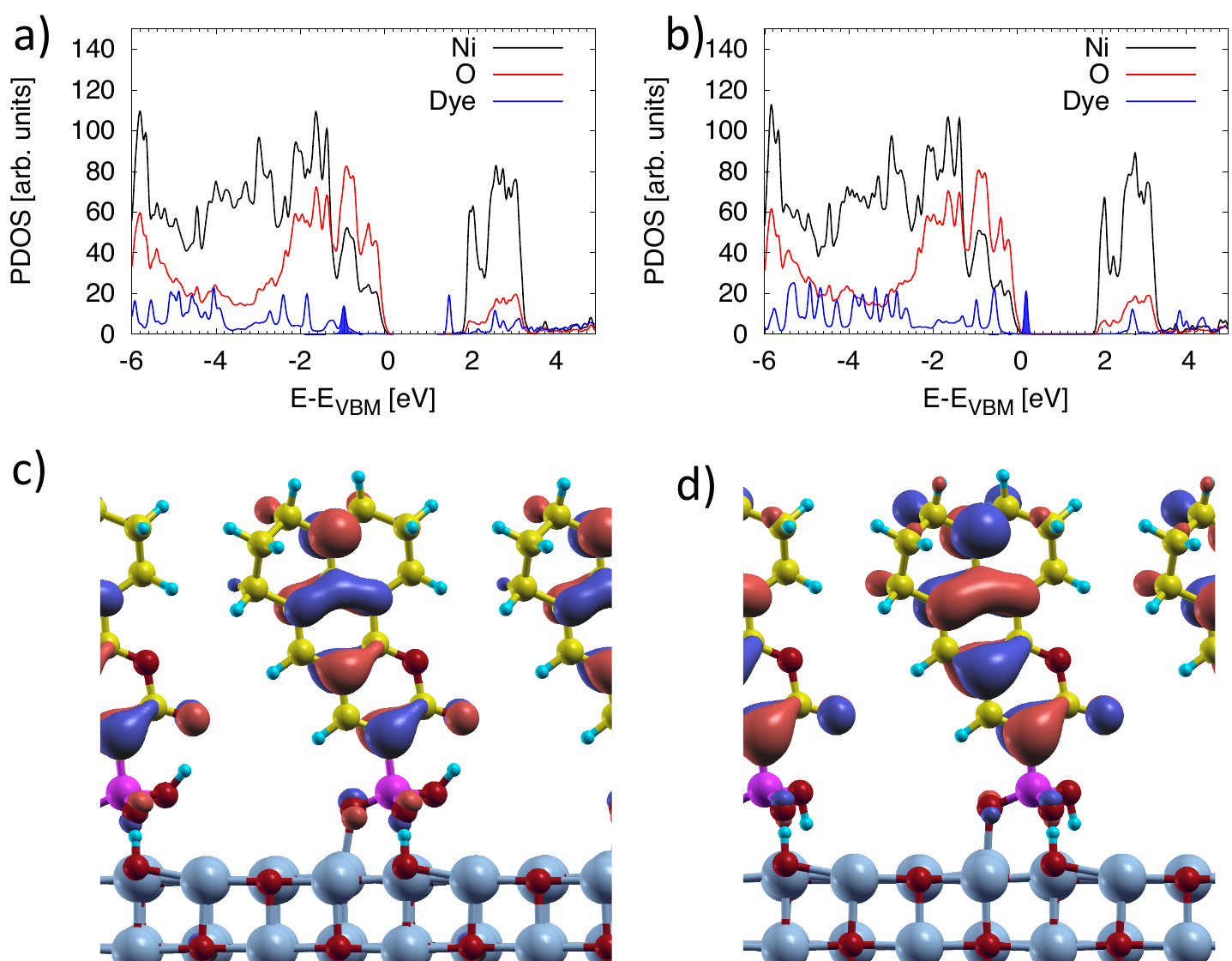}
\caption{\label{Fig_PDOS_asd} The C343/NiO(100) system in vacuum in the bidentate anchoring geometry, evaluated using the PBE+U approach. We show the PDOS of: (a) the H-up geometry; (b) H-down geometry. The blue shaded areas represent the projections on the HOMO of the isolated dye. The isosurfaces of the dye's HOMO Kohn-Sham state are shown in panel (c) for the H-up geometry and panel (d) the H-down geometry. In both cases the dye's HOMO is fully localized on the dye, with no significant weight on the NiO slab. 
Reprinted with permission from 
(Piccinin, S.; Rocca, D; Pastore, M. Role of Solvent in the Energy Level Alignment of Dye-Sensitized NiO Interfaces \emph{J. Phys. Chem. C}, \textbf{2017}, \emph{121}, 22286-22294). Copyright
(2017) American Chemical Society.
}      
\end{figure}

In line with previous work \cite{Munoz-Garcia2015} the bidentate binding was found to be the most stable anchoring 
mode of the CH$_3$-PO(OH)$_2$ group onto the NiO(100) surface with an adsorption energy for the  H-up and H-down C343 isomers adsorbed on a 6-layer 
surface slab of  $-$1.30 eV and $-$1.06 eV, respectively.  In addition to the  different adsorption energy, favoring the H-up configuration, 
the PDOS for the two systems in \ref{Fig_PDOS_asd} show that 
there is also a large influence on the position of the molecular orbitals of the dye with respect to the NiO levels. 
In particular the HOMO of the dye lies 1.1 eV below the valence band maximum (VBM) in the H-up configuration, while it is 0.1 eV above the VBM in the H-down geometry, 
the latter coinciding with the relative alignment predicted for the isolated components. 
The explanation for this lies in the fact that the different adsorption geometries result in 
different interfacial dipoles, which shift the electrostatic potential in the region of the dye relative to the potential of the NiO slab (the shift in the 
electrostatic potential perpendicular to the slab is exactly 1.2 eV in the region of the dye).\cite{Piccinin-2017} 
Being essentially decoupled from the NiO slab (see  plot of the corresponding Kohn-Sham state in Figure \ref{Fig_PDOS_asd} and the sharp peak of the dye's HOMO), 
the HOMO of the dye experiences a rigid shift in energy with respect to the 
valence band of NiO as the positively charged H atom of the anchoring group moves from the H-down to the H-up configuration. 

\subsection{Charge separation and electron injection rates}

The first and most important electron-transfer process
occurring at the dye/semiconductor interface is the charge injection from the photoexcited
dye to the CB of the substrate.\cite{Ponseca-2017} 
The injection quantum yield $\eta_{inj}$ (photons absorbed by the dye
that end uo in electron injected into the CB of the metal oxide)
is given by:
\begin{equation}\label{eqn:etainj}
\eta_{inj}=\frac{k_{inj}}{k_{inj}+k_0+k_q}
\end{equation}
where $k_0$ is the rate constant for radiative and nonradiative decay
of the excited state of stand-alone sensitizer and $k_q$ represents all
possible deactivation channels of the dye's excited state in the
cell different from electron injection.
The remarkable fortune obtained by ruthenium dyes in the DSC field
is at least partially ascribable to their high electron injection quantum
yields,\cite{ArdoChemSocRev2009,ListortiEnergyEnvironSci2011,
ListortiChemMater2011,
AndersonAnnuRevPhysChem2004,LobelloJPhysChemC2011,
KoopsJPhysChemC2010}
originating from the large values of the electron injection
rates ($\sim$100 ps) but also from the long lifetimes (from tens to
hundreds of nanoseconds) of the lowest-energy triplet state (mostly
responsible of the electreon injection) that strongly reduce the term $k_0$ in 
equation~\ref{eqn:etainj}.\cite{KoopsJPhysChemC2010}
In Ru(II) compounds, electron injection is also assisted by the small 
recombination rate constants as a consequence of an efficient
electron-hole separation and by their peculiar grafting
geometry, which adequately hinders the interfacial aggregation
phenomena.\cite{KoopsJPhysChemC2010} However, despite its record efficiencies, 
in the view of large-scale solar energy production, ruthenium suffers from serious drawbacks, 
which potentially limit its widespread applicability, mainly related to its toxicity and scarcity. 
This has motivated continuous research efforts to develop valuable alternatives, exploiting 
earth-abundant, less expensive and environmentally friendly d-block metal (for example iron and copper) 
complexes.\cite{Ferrere-1998,Yang-2000,Dixon-2015,Bessho-2008,Housecroft-2015} 
Despite an intense absoprtion by the metal-to-ligand charge transfer (MLCT) states, however, conventional Fe(II)-polypyridyl complexes are, unfortunately, 
characterized by an ultrafast (ca. 100 fs) deactivation to low-lying Metal Centered (MC) states, 
proceeding via the triplet $^3$MC and ultimately populating the quintuplet $^5$T2 state,\cite{McCusker-1993,Monat-2000,Cammarata-2014} 
which impedes electron injection into the sensitized semiconductor.\cite{VanKuiken-2016}
Recent breakthroughs have been obtained by the development of Fe(II) complexes 
with strongly electron $\sigma$-donating N-heterocyclic carbene (NHC) ligands,\cite{Liu-2013,Fredin-2014,Liu-2016,Duchanois-2014,Harlang-2015,Zimmer-2018}
reaching record lifetimes of hundreds of picoseconds.\cite{Chabera-2017,Chabera-2018}
On the other hand, for organic dyes the
thermalization to the triplet state is much slower than the
injection and emission processes, and hence the system 
transfers the photo-excited electron directly from the lowest singlet state. Even though 
the injection process from the singlet is faster compared to that from
the triplet state, the singlet lifetime can be abuot 20-fold
shorter, opening the way to a considerably larger competition
between the injection process and the radiative decay, yielding
to lower quantum yields.\cite{KoopsJPhysChemC2010}
Moreover, organic chromophores are usually characterized by faster recombination of the injected electron
with the oxidized dye when compared to metal complexes,
possibly originated by a generally closer localization of the dye
highest occupied molecular orbital (HOMO) to the TiO$_2$
surface,\cite{KoopsJPhysChemC2010} also boosted by particular grafting configurations.\\

The theoretical
description of excited-state electron injection into wide-bandgap
semiconductors was first proposed by Marcus and 
Gerischer.\cite{GerischerTopCurrChem1976,
GerischerPhotochemPhotobiol1972,MarcusJChemPhys1965}
Under the weak coupling regime, the rate constant
($k_{inj}$) for the electron transfer from a single dye's $d$ state to
the many acceptor states $k$ of the metal oxide can be
obtained using the Fermi Golden Rule as:
\begin{equation}\label{eqn:kinj}
k_{inj}=\frac{2\pi}{\hbar}\displaystyle\sum_{k}|V_{dk}|^2\rho(\varepsilon_k)
\end{equation}
where $\rho(\varepsilon_k)$ is the density of semiconductor acceptor 
states (DOS) at the energy $\varepsilon_k$, $V_{dk}$ the electronic coupling
between the (diabatic) donor $d$ state and the $k$th acceptor
state in TiO$_2$, and $\hbar$ Planck's constant. 
A rapid estimate of the electron injection times can be attained by
resorting to the Newns-Anderson model.\cite{MuscatProgSurfSci1978,
CohenTannoudjiBook1977,PerssonJChemTheoryComput2006,
LundqvistJPhysChemB2006,PerssonJPhysChemB2005,PerssonIntJQuantumChem2002} 
This approach provides a fast assessment of
the effects of the adsorption on the molecule electronic levels
(generally only the dye LUMO is analyzed), characterizing them
in terms of an energy shift relative to the free system and a lifetime
broadening, $\hbar\Gamma$, defined by a
Lorentzian distribution resulting from the decay of the dye
excited state, approximated by the dye LUMO, coupled to
the continuum of the TiO$_2$ CB states\cite{Pastore-2015,AnselmiPhysChemChemPhys2012,Umari-2013,Calbo-2014,
RoncaJPhysChemC2014Prot,Lasser-2015} 
To obtaine these
quantities one hase to calculate the projected density of states
(PDOS) relative to the dye's LUMO in the dye-TiO$_2$ complex. When the
system's molecular orbitals in a certain atomic basis are
expanded, the contributions $p_i$ to the dye's LUMO PDOS are
defined by the relation:
\begin{equation}
p_i=\displaystyle\frac{\sum_j^{A\in dye}(c_{ij}^A)^2}{\sum_j^n(c_{ij}^A)^2}
\end{equation}
where $c_{ij}^A$ are the expansion coefficients of the molecular
orbitals of the complex on the basis function of atom $A$ belonging to the dye.

The center of this distribution gives the energy of the
dye's LUMO grafted on TiO$_2$, $E_{LUMO}(ads)$, and it can be
obtained as:
\begin{equation}\label{eqn:ELUMO}
E_{LUMO}(ads)=\displaystyle\sum_i p_i\varepsilon_i.
\end{equation}

On the other hand, the width of the LUMO broadening can be
estimated as a mean deviation of a distribution centered at the
$E_{LUMO}(ads)$ energy value through the equation:
\begin{equation}\label{eqn:Gamma}
\hbar\Gamma=\displaystyle\sum_i p_i|\varepsilon_i-E_{LUMO}(ads)|.
\end{equation}

The Newns-Anderson model, despite its simplicity and effectiveness, 
does not allow the direct calculation of the
coupling between the dye and semiconductor states, leading
directly to the injection time. To do that, one can 
use the model proposed by Thoss and co-workers.\cite{KondovJPhysChemC2007,LiJChemPhys2012}
This approach, based on the localization of the molecular
orbitals of the complex on the donor ($d$) and acceptor ($a$)
species, gives the diabatic semiconducductor 
DOS and the diabatic dye's LUMO, as well as the explicit coupling between the dye's LUMO and the manifold of the semiconductor acceptors states. 
The product between the square of the electronic coupling elements, $|V_{dk}|^2$, and the TiO$_2$ 
density of states, $\rho(\varepsilon_k)$, 
defines the electron transfer probability distribution, $\Gamma(\varepsilon_k)$. 
Finally, electron transfer times can be evaluated by $\tau$(fs) = 658/$\Gamma$(meV).

In the following we will review the application of the two above mentioned approaches  
to TiO$_2$ substrates (bares and protonates) sensitized by organic dyes having different electronic structure and excited states properties.\cite{RoncaJPhysChemC2014Prot}

\subsubsection{Role of the Dye Molecular Structure and of the 
CB Position on Injection Rates}

\begin{figure}[!ht]
\centering
\includegraphics[width=8cm]{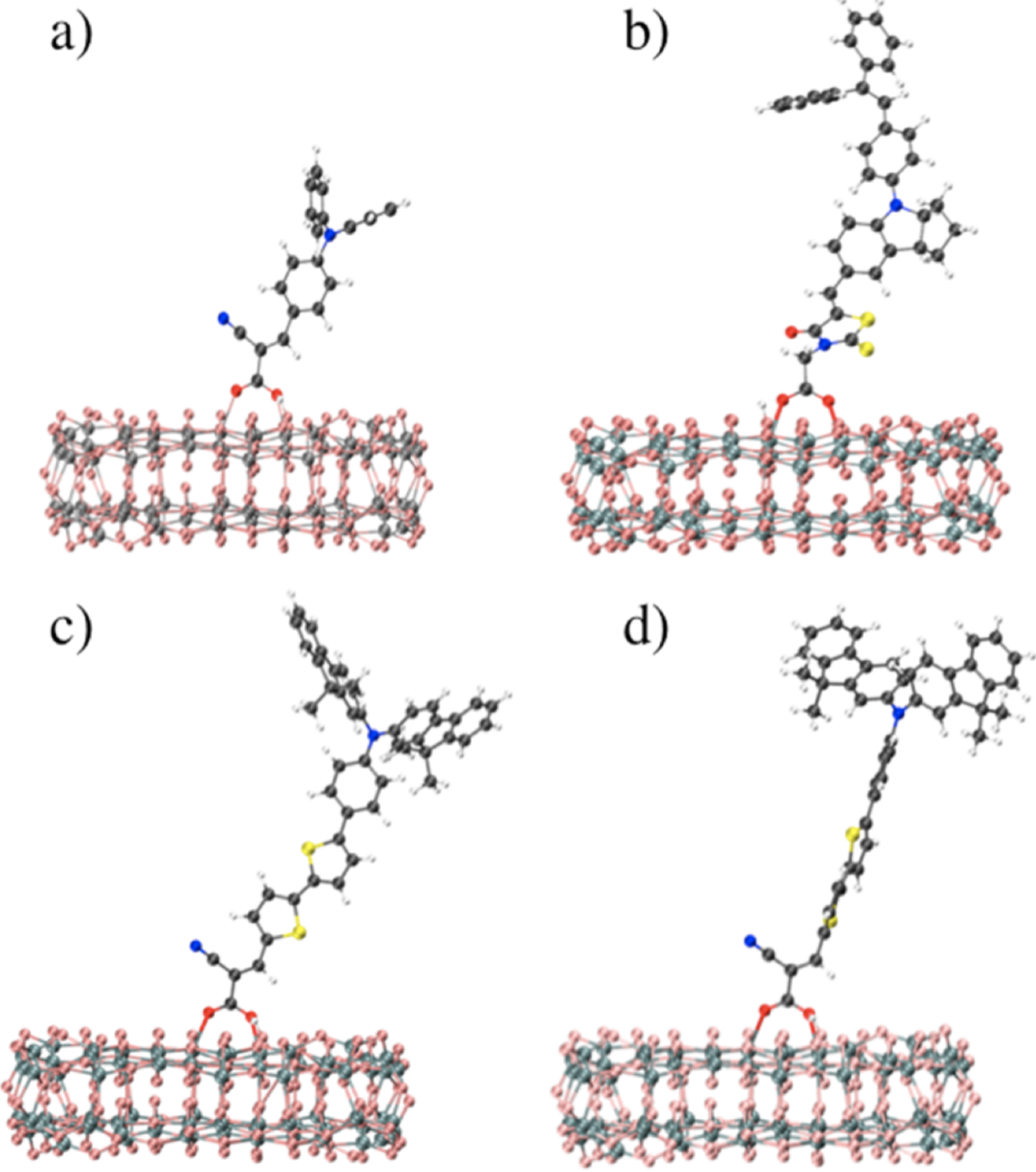}
\caption{\label{fig:dyes_TiO2_struct}\small Optimized geometries 
of the L0 (a), D102 (b), JK2 (c), and
JK2-T (d) dyes adsorbed onto the (TiO$_2$)$_{82}$ cluster in bridged bidentate
(BB) anchoring geometry. 
Reprinted with permission from 
(Ronca, E.; Marotta, G.; Pastore, M; De Angelis, F. Effect of Sensitizer Structure and TiO$_2$ Protonation on Charge
Generation in Dye-Sensitized Solar Cells \emph{J. Phys. Chem. C}, \textbf{2014}, \emph{118}, 16927-16940). Copyright
(2014) American Chemical Society.
}
\end{figure}

Two popular cyanoacrylic based
conjugated dyes (Figure \ref{fig:dyes_TiO2_struct}) 
differing for the extension of the $\pi$ bridge, namely, 
L0,\cite{HagfeldtBook2010,HagbergJOrgChem2007,
PastorePhysChemChemPhys2012} and JK2\cite{PastoreJPhysChemC2010,KimJAmChemSoc2006,
ChenNanoLett2009,PastoreJPhysChemC2013} will be examined.\cite{RoncaJPhysChemC2014Prot}
For the latter we also considered a different configuration, obtained by
twisting of 90$^{\circ}$ the C-C bond linking the cyanoacrylic
anchoring and thiophene units (JK2-T, see below), to assess
the effects of breaking of the electron conjugation. Then, to enrich our set, also 
an indoline-based dye, D102, was included.\cite{PastoreJPhysChemC2010}
As shown in Figure \ref{fig:dyes_TiO2_struct}, the sensitizers have been adsorbed onto the TiO$_2$
surface by the carboxylic group in the BB configuration, 
upon deprotonation and H transfer to the surface.\cite{PastoreACSNano2010,
PastorePhysChemChemPhys2012}

\begin{figure}[!ht]
\centering
\includegraphics[angle=90,width=9cm]{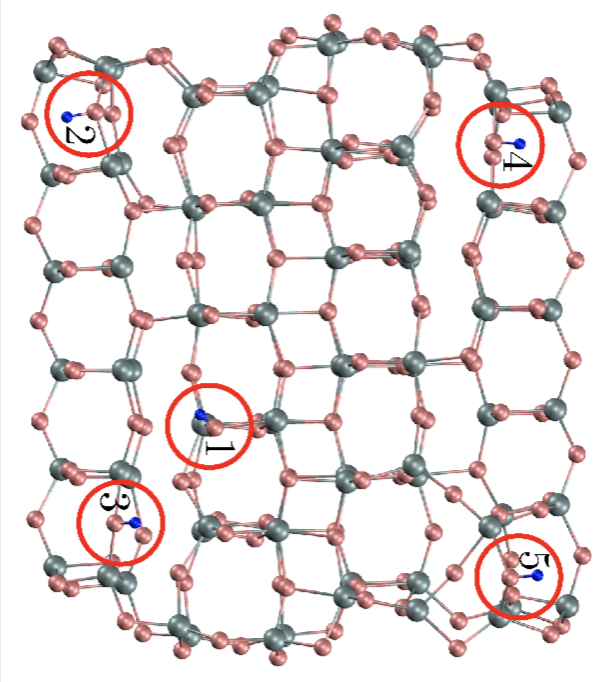}
\caption{\label{fig:TiO2_prot}\small Scheme of the adsorption position of protons on the
(TiO$_2$)$_{82}$ cluster. 
Reprinted with permission from 
(Ronca, E.; Marotta, G.; Pastore, M; De Angelis, F. Effect of Sensitizer Structure and TiO$_2$ Protonation on Charge
Generation in Dye-Sensitized Solar Cells \emph{J. Phys. Chem. C}, \textbf{2014}, \emph{118}, 16927-16940). Copyright
(2014) American Chemical Society.
}
\end{figure}

The TiO$_2$ surface has been modeled by a
(TiO$_2$)$_{82}$ cluster,\cite{DeAngelisJPhysChemC2010} 
obtained by an appropriate ``cut'' of
an anatase slab exposing the majority (101) 
surface.\cite{VittadiniPhysRevLett1998}
To model  
surface protonation we adsorbed on the TiO$_2$ cluster surface an
increasing number of protons, as shown in Figure \ref{fig:TiO2_prot}. 
The first
proton was placed at the center of the cluster, close to the
dye, while the other four H$^+$'s have
been progressively adsorbed at the cluster corners.

The dye@(TiO$_2$)$_{82}$ equilibrium geometries were
obtained in \emph{vacuo}
with the ADF program package\cite{ADF2001} employing the PBE exchange
correlation functional with the TZP (DZP) basis set for Ti
(H,C,N,O,S) atoms.
The Projected Density of States (PDOS) relative to the dye's LUMO were
obtained by single-point calculations performed at
the DFT/B3LYP level of theory using the SVP basis 
set  on
the ADF-optimized geometries. Solvation effects (acetonitrile)
were modeled by means of the 
C-PCM\cite{cpcm2}
method implemented in the Gaussian 09 (G09) suite.\cite{g09} 

\begin{figure}[!ht]
\centering
\includegraphics[width=12cm]{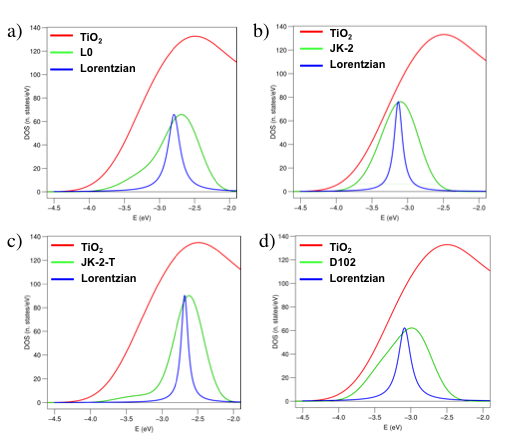}
\caption{\label{fig:DOS_1H}\small PDOS relative to the sensitizer and to TiO$_2$ 
and corresponding Lorentzian distribution of the dye's LUMO for L0 (a), JK2 (b), JK2-T (c) and
D102 (d) in their neutral form (1H). The TiO$_2$ PDOS have been 
normalized; those corresponding to the sensitizer have been normalized and multiplied
by 50; and the Lorentzian distributions have been normalized on the dye's PDOS maximum. Figure adapted from Ref. \cite{RoncaJPhysChemC2014Prot}}
\end{figure}

The results for the investigated
systems in the neutral monoprotonated form (hereafter termed
1H), starting with the Newns-Anderson analysis of the dye-
TiO$_2$ complexes, are reported in Figure~\ref{fig:DOS_1H}, 
where we also report the corresponding Lorentzian distribution and the PDOS
relative to the semiconductor in the presence of the dye.

The TiO$_2$ PDOS have basically the same profile of
that of the corresponding bare TiO$_2$ slab, apart
from a small upshift (ca. 0.1-0.2 eV) induced by the adsorbed dye (see the
CB edge energetic values for the isolated semiconductor (E$_{CB}$)
and for the TiO$_2$ in the complex (E$_{CB-complex}$) in 
Table~\ref{tab:data_1H}). 

\begin{table}[!htp]
\caption{\label{tab:data_1H}\small 
Calculated CB  of the bare TiO$_2$ at the geometry 
It has in the complex (E$_{CB}$), Energy of the
LUMO of the dye anchored on the TiO$_2$ (E$_{LUMO}$), 
CB of the TiO$_2$ when the dye is adsorbed
(E$_{CB-complex}$) and Lorentzian Broadening ($\hbar\Gamma$) 
for the investigated systems in 1H Form. Data from Ref. \cite{RoncaJPhysChemC2014Prot}}
\vspace{0.5cm}
\centering
\begin{tabular}{lccccc}
Dye &E$_{CB}$ (eV) &E$_{HOMO-complex}$ (eV) &E$_{LUMO}$ (eV) &E$_{CB-complex}$ (eV)  &$\hbar\Gamma$ (eV)\\
\hline
L0 &-3.999 &-5.622 &-2.793 &-3.845  &0.210\\
JK2 &-3.999 &-5.105 &-3.131 &-3.844 &0.141\\
JK2-T &-4.000 &-5.013 &-3.087 &-3.859 &0.209\\
D102 &-4.000 &-5.191 &-2.683 &-3.839  &0.116\\
\hline
\end{tabular}
\end{table}
These energy shifts come out from the combined effect of the electrostatic 
potential generated by
the dye on the TiO$_2$ surface and of the
charge moving from the dye to the TiO$_2$ by virtue of the 
dye-semiconductor interactions, as detailed above.\cite{RoncaEnergyEnvironSci2013} 
On the other hand, when the dye is bound
to the semiconductor, the LUMO tends to have significant
spread over a large number of eigenvalues, as shown by
the sizable broadenings presented by the dye's LUMO PDOS
(green lines in Figure~\ref{fig:DOS_1H}). 
By the Newns-Anderson
model, we can calculate the energy of the LUMO for 
the adsorbed dye and its relative shift with respect to the value in the isolated dye, 
(equation~\ref{eqn:ELUMO}), the energetic Lorentzian distribution (Figure \ref{fig:DOS_1H}) and the corresponding broadening (equation~\ref{eqn:Gamma}). 

Starting our discussion from the system containing the L0 dye
(\ref{fig:DOS_1H}a), we observe a sizeable broadening
amounting to 0.21 eV, which delivers an electron injection time
of about 3 fs, in qualitative agreement
with the theoretical and experimental reference data. In
particular, this value is comparable
with those evaluated by the use of different and more refined theoretical methods
(2-10 fs, see Ref.\cite{JonesPhysChemChemPhys2010} and 
Ref.\cite{MartsinovichJPhysChemC2011}), 
for similar organic sensitizers and is
certainly less than 200 fs as expected from experimental 
measurements.\cite{WibergJPhysChemC2009,
WibergJPhysChemB2010}

Turning now to the JK2 dye, the plots in Figure~\ref{fig:DOS_1H}b show that
the sensitizer's LUMO PDOS is centered at lower
energies when compared to that of L0, in line with the LUMO energies of the isolated dyes
(-3.00 eV and -2.54 eV for JK2 and L0 respectively). In contrast to what was obtained for L0,
the JK2 adsorbed
LUMO is mainly localized on the dye states, as
evidenced by the largely smaller value of the energetic
broadening (0.14 eV), resulting in a 40\% increased electron
injection time with respect to that computed for L0,
consistent with the reference data from
the literature.\cite{WibergJPhysChemC2009,JonesPhysChemChemPhys2010,
MartsinovichJPhysChemC2011,WibergJPhysChemB2010}
 Despite this, JK2 shows good electron injection
properties, as suggested by the high efficiency ($\sim$8\%) obtained
in cells built using this sensitizer.\cite{KimJAmChemSoc2006}
When dealing with the injection rate, two main variables, related to the dye's molecular structure, 
have to be considered:\cite{JonesPhysChemChemPhys2010} 
the electron
injection energy (E$_{LUMO}$), which dictates the driving force, and the percentage of LUMO localized
on the anchoring moiety, which dictates the electronic coupling. 
We can attribute a part of the variations observed in the injection rates to the above discussed
shift between the LUMO energies of the two dyes in
both the adsorbed and standalone forms. In particular, by a
comparison of the Lorentzian distributions with the semiconductor
PDOSs, we notice that L0 presents the LUMO in an
energetic region where the TiO$_2$ states are more dense, while the
JK2 LUMO is energetically localized close to the CB edge, where
there are few semiconductor empty states available for the
mixing. Turning to the localization of the dye's LUMO, 
the percentage of
the LUMO on the anchoring moiety atoms can be obtained by
summing the $p_i$ contributions of the LUMO relative to the atoms
of the cyanoacrylic group. A similar procedure can be applied also
to the dye adsorbed on the semiconductor by performing the
sum on a number of complex states containing a unitary
contribution of the sensitizer. As expected, L0, in
its free protonated form, has about 31\% of the LUMO localized
in the anchoring moiety while for JK2 this state is more
distributed along the molecule and shows a percentage of 24\% on
the cyanoacrylic group; if we consider the dyes in the complex
with the TiO$_2$, the two percentages increase, but a
comparable difference between the two systems still remains
(66\% for L0 and 51\% for JK2). To further elucidate this effect, we investigated, the JK2-T dye, 
where we twisted by 90$^{\circ}$ the C-C bond
connecting the cyanoacrylic anchoring moiety to the first
thiophene ring of the $\pi$-spacer; the corresponding PDOSs are
reported in Figure~\ref{fig:DOS_1H}c. It is worth noting that while the dye's
LUMO energy is almost unaffected by the twisting of the bond
(the energy change is only 0.05 eV), the broadening of the
LUMO distribution is strongly modified (Figure~\ref{fig:DOS_1H}c). By the
analysis of the corresponding Lorentzian distribution, we
quantified the increase in the broadening, amounting to about
0.07 eV. From these values we then evaluated the injection time
for the JK2-T dye, obtaining a value of 3 fs. As the breaking of the
conjugation has a negligible effect on the LUMO energy, this
large increase in the injection rate can be completely attributed to
the increase of the LUMO localization on the anchoring region.
As a matter of fact, passing from the planar to the twisted JK2
configuration, the LUMO localization on the cyanoacrylic
moiety increases by about 18\% and 35\% in dye in its free and
adsorbed form, respectively.\cite{RoncaJPhysChemC2014Prot}

We conclude with the D102
dye, attached to TiO$_2$ by a nonconjugated 
rhodanine-3-acetic anchoring moiety\cite{PastoreACSNano2010} 
(Figure~\ref{fig:dyes_TiO2_struct}b). 
The LUMO PDOS maximum (Figure \ref{fig:DOS_1H}d) is
energetically localized at higher energies with respect to that of
the other dyes, and a small stabilization
(0.07 eV) of the lowest-energy unoccupied state between the free
(-2.61 eV) and adsorbed system is obtained. This is indicative of a small
coupling between the sensitizer and the semiconductor states. 
Indeed the calculated energetic broadening is 0.12 eV, yielding an
increased injection time of $\sim$6 fs. 
The
LUMO component on the carboxylic unit is negligible, as
for D102 anchored on TiO$_2$, the LUMO localization on the carboxylic group 
amounts to 12\%, significantly smaller than
those relative to the other dyes (L0@TiO$_2$, 22\%; JK2@TiO$_2$, 
17\%; JK2-T@TiO$_2$, 26\%). Despite this very low localization
of the injecting orbital on the anchoring unit, the charge transfer
rate, remains in the ultrafast timescale
(within 100 fs). 
These relatively fast injection kinetics can be explained
considering the LUMO energetic position, localized in a dense
region of TiO$_2$ states ($\sim$126 states/eV for the bare semiconductor
and $\sim$130 states/eV for the TiO$_2$ in the presence of
the sensitizer, Table~\ref{tab:data_1H}).

\subsubsection{TiO$_2$ Protonation: Energy Level Shifts and 
Dye-Semiconductor Electronic Coupling}

How does TiO$_2$ protonation modify the relative
dye/semiconductor energy level alignment and the electronic
coupling? The effect on the injection rates upon surface protonation can be basically
attributed to two interplaying factors: (i) the dependence of the
TiO$_2$ DOS and of the dye's LUMO position on the number of H$^+$
adsorbed on the surface and (ii) the variation of the coupling
matrix elements between the sensitizer and the semiconductor.
Here we go beyond the Newns-Anderson model, by directly calculating the electronic coupling 
between the localized dye's and semiconductor's states. 

\begin{figure}[!ht]
\centering
\includegraphics[width=14cm]{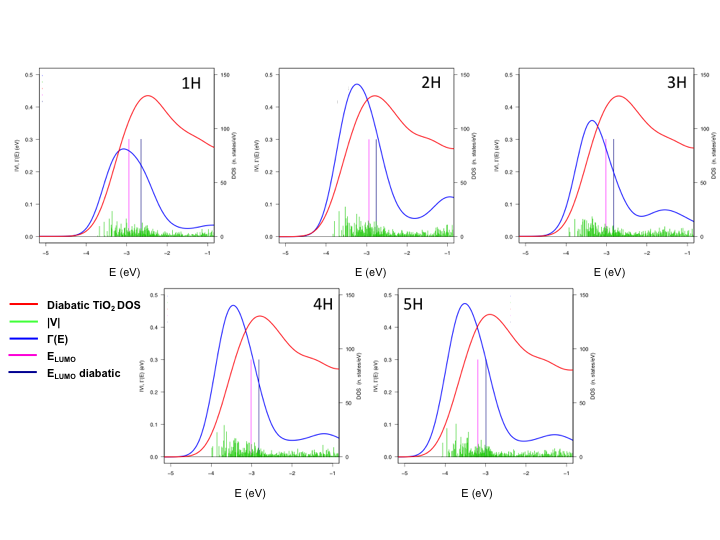}
\caption{\label{fig:DOS_L0_protvar}\small Absolute value of the 
coupling matrix elements ($V_{dk}$) between the dye LUMO and the TiO$_2$ 
virtual states (green sticks), injection broadening
$\Gamma$(E) function (blue line), and DOS relative to 
the virtual localized (diabatic) orbital of the 
semiconductor (red line) for L0@TiO$_2$ in different
protonation forms as a function of the energy (E). 
For each system the magenta and blue vertical lines 
represent the energetic position of the adsorbed
dye LUMOs evaluated as the maximum of the 
Newns-Anderson Lorentzian distribution 
and as the energy of the localized orbital more similar to the
sensitizer's first virtual state, respectively. 
The Gaussian broadening used to reproduce 
both the TiO$_2$ DOS and the $\Gamma$(E) function is equal to 0.3 eV. Figure adapted from Ref. \cite{RoncaJPhysChemC2014Prot}}
\end{figure}

\begin{figure}[!ht]
\centering
\includegraphics[width=12cm]{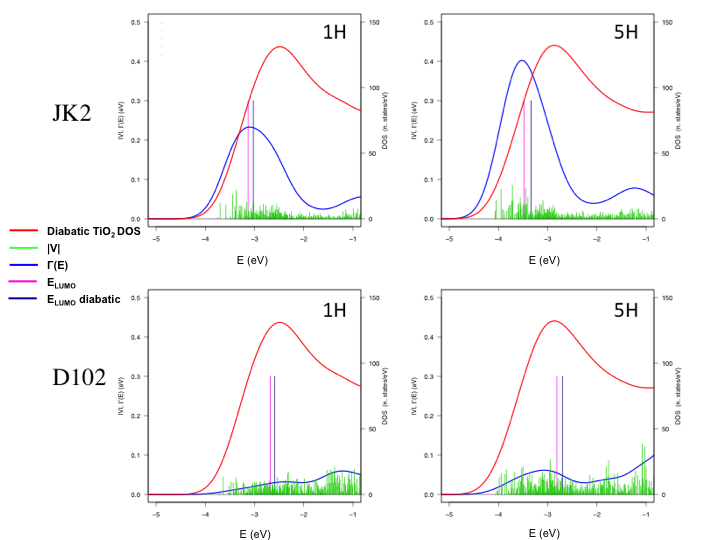}
\caption{\label{fig:DOS_JK2_D102_protvar}\small Absolute value of the coupling 
matrix elements ($V_{dk}$) between the dye LUMO and the TiO$_2$ 
virtual states (green sticks), injection broadening
$\Gamma$(E) function (blue line), and DOS relative to the 
virtual localized (diabatic) orbital of the semiconductor 
(red line) for the JK2@TiO$_2$ and D102@TiO$_2$ systems in 
the 1H (left panels) and 5H (right panels) protonation 
form as a function of the energy (E). For each system, the magenta and blue
vertical lines represent the energetic position of the adsorbed 
dye LUMOs evaluated as the maximum of the Newns-Anderson Lorentzian distribution
and as the energy of the localized orbital more similar to 
the sensitizer first virtual state, respectively. 
The Gaussian broadening used to reproduce both
the TiO$_2$ DOS and the $\Gamma$(E) function is equal 
to 0.3 eV. Note that in the case of the D102 dye 
(bottom panels) both the coupling matrix elements ($V_{dk}$)
and the $\Gamma$(E) function were multiplied by 10. Figure adapted from Ref.\cite{RoncaJPhysChemC2014Prot}
}
\end{figure}

The results for the L0@TiO$_2$ system in all the modeled
protonation forms (from 1H to 5H) are reported in 
Figure~\ref{fig:DOS_L0_protvar},
while those for the JK2 and D102 dyes in both 1H and 5H forms
investigated are displayed in Figure~\ref{fig:DOS_JK2_D102_protvar}. 

In Table~\ref{tab:data_protvar} we collect some relevant
quantities extracted upon the diabatization step: the energy of
the diabatic TiO$_2$ CB; the dye's HOMO and LUMO, with the
associated HOMO-LUMO gap; and the injection function
$\Gamma$(E) evaluated at the energy of both diabatic and Lorentzian
(NA) LUMO, which is also reported for the sake of
completeness.
\begin{table}[!htp]
\caption{\label{tab:data_protvar}\small Calculated Energies (Electronvolts) 
of Diabatic TiO$_2$ Conduction Band (CB$_{DIA}$), Dye's HOMO and LUMO and 
HOMO-LUMO Gap (GAP$_{DIA}$), Injection Function ($\Gamma$(E)) 
Evaluated at the Energy of Both Diabatic ($\Gamma^{DIA}$, LUMO$_{DIA}$) and Lorentzian
($\Gamma^{DIA}$, LUMO$_{NA}$) LUMO, and Lorentzian LUMO (LUMO$_{NA}$). 
Reprinted with permission from 
(Ronca, E.; Marotta, G.; Pastore, M; De Angelis, F. Effect of Sensitizer Structure and TiO$_2$ Protonation on Charge
Generation in Dye-Sensitized Solar Cells \emph{J. Phys. Chem. C}, \textbf{2014}, \emph{118}, 16927-16940). Copyright
(2014) American Chemical Society.
}
\vspace{0.2cm}
\centering
\footnotesize
\begin{tabular}{lccccccc}
H$^+$ &CB$_{DIA}$ &LUMO$_{DIA}$ &HOMO$_{DIA}$ &GAP$_{DIA}$ &$\Gamma^{DIA}$(LUMO$_{DIA}$) &LUMO$_{NA}$ &$\Gamma^{DIA}$(LUMO$_{NA}$)\\
\hline
\multicolumn{8}{c}{L0}\\
1H &-3.830 &-2.643 &-5.603 &2.959 &0.220 &-2.793 &0.247\\
2H &-4.132 &-2.762 &-5.685 &2.923 &0.327 &-2.945 &0.410\\
3H &-4.042 &-2.826 &-5.721 &2.895 &0.199 &-3.012 &0.271\\
4H &-4.132 &-2.921 &-5.783 &2.862 &0.279 &-3.120 &0.377\\
5H &-4.229 &-2.991 &-5.836 &2.845 &0.302 &-3.190 &0.392\\
\multicolumn{8}{c}{JK2}\\
1H &-3.846 &-3.025 &-5.100 &2.074 &0.231 &-3.131 &0.233\\
5H &-4.220 &-3.337 &-5.214 &1.877 &0.375 &-3.484 &0.401\\
\multicolumn{8}{c}{D102}\\
1H &-3.999 &-2.593 &-5.188 &2.596 &0.003 &-2.683 &0.003\\
5H &-4.391 &-2.699 &-5.188 &2.489 &0.005 &-2.815 &0.006\\
\hline
\end{tabular}
\end{table}

Let us start by discussing the case of L0: as is apparent from the
plots in Figure~\ref{fig:DOS_L0_protvar}, we observe a very
regular increase in the electronic coupling at lower energies as the number of protons on the surface increases,
with the only exception being the 2H case, which shows an
exaggerated increment in the low energy
region, possibly due to the fact that the only positive
charge is located at one of the cluster corners. As higher energies
are considered, the coupling values progressively tend to
decrease. The $\Gamma$(E) functions (obtained by the product of the square of the
electronic coupling and the DOS) show a pronounced maximum
in correspondence of the TiO$_2$ CB edge region, where the coupling is higher. 
The progressive increasing of the protonation degree yields a sizable energetic
shift of the $\Gamma$ distribution maximum, which is clearly related to
semiconductor CB edge lowering (0.1 eV per proton added). 
By extracting the value of $\Gamma$(E) at the sensitizer's LUMO
energy, one can estimate the charge injection rate from the dye to
the semiconductor. It is worthwhile to stress that the position of
the dye LUMO can be evaluated in two different ways: in the
adiabatic picture, as the energy of the Newns-Anderson
Lorentzian distribution maximum (LUMO$_{NA}$, magenta sticks
in Figures~\ref{fig:DOS_L0_protvar} and \ref{fig:DOS_JK2_D102_protvar}) 
and, in the diabatic framework, as the energy
of the localized orbital corresponding to the dye LUMO (dark
blue sticks in Figures~\ref{fig:DOS_L0_protvar} 
and \ref{fig:DOS_JK2_D102_protvar}). We
then extracted the $\Gamma$ function intensity in the dye LUMO range,
obtaining a value of about 0.22-0.25 eV (corresponding to about
3 fs) for the 1H form and 0.30-0.39 eV ($\sim$2 fs) for the 5H case.
Similar conclusions can be drawn also for the system containing
the JK2 dye (upper panels in 
Figure~\ref{fig:DOS_JK2_D102_protvar}). Comparing the $\Gamma$(E) curves with those
of the L0 dye, we can immediately notice a lowering of the
injection rates (about 0.1 eV at the maxima energetic position) in
the whole energy range and for both the protonation, imputable
to the smaller spatial localization of the LUMO on the anchoring
unit found in JK2 compared to L0. This trend is consistent with
the results provided by the Newns-Anderson model and
discussed in the previous section. However, by extracting the
$\Gamma$(E) function in the dye LUMO range, we obtain injection rate
values of 0.23 eV for the 1H form and 0.37-0.40 eV for the 5H
case, which result in values somewhat larger than those calculated
for the corresponding L0 systems. This is clearly the result of a
more favorable energetic position of the dye's LUMO, which,
lying at lower energies, can inject into the CB states for which the
electronic coupling is higher (i.e., the LUMO position falls in the
region where the $\Gamma$(E) function shows the maximum).

Turning now to the nonconjugated D102 dye, comparing the
injection rates in Table~\ref{tab:data_protvar} 
with those reported in Table~\ref{tab:data_1H}, we
observe a substantial disagreement between the results provided
by the two approaches. In fact, by considering the explicit
coupling between the LUMO and the CB states, we obtain
electronic coupling values of 0.003 (1H) and 0.005 eV (5H), whereas the
values obtained by the Lorentzian broadening are 0.12 (1H) and
0.16 eV (5H). While the diabatic results are clearly consistent
with the not conjugated structure of the dye and with the
negligible localization of the dye's LUMO on the carboxylate
anchoring group, the broadening estimated in the 
Newns-Anderson model by means of a Mulliken population analysis is
evidently sensitive to the basis set quality choice, a well-know
drawback of this kind of electron population scheme, where the
electron density is projected onto the basis 
set.\cite{CioslowskiJAmChemSoc1989}

Final remarks concern the effect of surface protonation 
on the dye and TiO$_2$ energy levels: comparing the CB and LUMO energy levels for
the different dyes, reported in Table~\ref{tab:data_protvar}, 
one can notice that while
the extent of the CB$_{DIA}$ down-shift as the number of protons
increases does not depend on the particular dye adsorbed,
different dyes show different down-shifts of the LUMO going
from the 1H to 5H forms. While the stabilization of the LUMO
(both diabatic and NA) is in the range 0.35-0.39 eV, comparable
to the ca. 0.40 eV shift of the CB edge, we observe a sizeably
smaller stabilization of the D102 LUMO, amounting to about 0.1
eV. This is clearly the consequence of the different electronic
structure of the interface: the energy levels of the conjugated dyes
(L0 and JK2) are more strongly perturbed by the surface
protonation, and the stabilization of the TiO$_2$ CB edge translates
into a similar stabilization of the dye's LUMO, while the LUMO
of the uncoupled D102 dye is practically unaffected by what
happens to the TiO$_2$ DOS. As a consequence of the progressive
charge accumulation on the surface, also the dye's HOMO turns
out to be stabilized in the case of the conjugated dyes: in the
shorter L0 system, where the ground-state electronic charge can
be more effectively transferred to the protonated 
TiO$_2$,\cite{RoncaEnergyEnvironSci2013} the
down-shift amounts to 0.23 eV, whereas in the longer JK2 dye,
the HOMO stabilization is about 0.1 eV. For the same reasons,
the D102 HOMO is practically unchanged upon TiO$_2$
protonation. The different HOMO and LUMO energy shifts
with surface protonation lead to an overall reduction of the dye's
HOMO-LUMO gap (see Table~\ref{tab:data_protvar}) with an 
increase in the number of adsorbed protons.
The entity of such HOMO-LUMO gap variation is 
clearly strongly dependent on the electronic
structure of the dye. 

Summarizing, high injection rates come out from the combination of high electronic coupling values
of the dye's LUMO with the TiO$_2$ CB states and high density of acceptor states. Therefore, maximizing the
LUMO's localization on the anchoring unit, favoring the electronic coupling, and shifting its 
position to relatively higher energies, where the TiO$_2$
DOS is denser, appear as optimal design rules to increase the electron injection efficiency.
Concerning the substrate protonation, on the other side, the present results 
demonstrate that the presence of protons on the TiO$_2$ surface, 
while shifting of both the semiconductor CB and dye's LUMO energetic positions, 
increases the injection rate mainly because of the increase in the coupling matrix elements, 
in particular with the substrate states in the lower energy region.

\section{Conclusion}
In the first part of this book chapter we revisited the theoretical elements necessary to elaborate qualitative and quantitative strategies for probing the structure, nature and locality of molecular exciton from excited states quantum-chemical calculations. For the qualitative analyses we exposed how the direct manipulation of quantum state charge densities can provide density functions leading to a clear picture of the transition. We also showed that as an alternative to the direct transposition of density matrices into the Euclidean space, one can also perform some algebraic operations on these matrices in order to obtain the hole/particle density matrices, called detachment/attachment density matrices. The direct inspection of the canonical orbitals was then exposed, as well as the possibility to derive two different orbitals sets (natural difference orbitals, natural transition orbitals) for excited states composed as a linear combination of singly-excited Slater determinants obtained from a single reference ground state wave function. In the special case of RPA, TDHF and TDDFT methods we also gave the definition of some transition orbitals (the projected natural transition orbitals and the canonical transition orbitals). On the other hand, quantitative analyses can also be classified according to the type of information provided: we exposed fragment-based analyses that assess the changes specific to a molecular fragment, global strategies which take into account the whole electronic system to provide an information related to the behavior of the entire molecule upon light absorption. Quantitative insights related to the statistical properties of the hole/particle distribution, or to its entanglement for example, were also reported. In the second part we, instead, discuss how the different electronic structure of the dye and dye-anchored metal oxide substrates directly influences the energy levels line-up at the interface, as well as the ground and excited state charge distribution and finally the electronic coupling and electron injection rates. In the case of electronically conjugated 
dye-TiO$_2$ interface, the substrate conduction band is up shifted, yielding higher V$_{oc}$, by a 
combined electrostatic (about 40\%) and ground state dye$\to$semiconductor charge transfer (about 60\%) effect. On the other hand, in the case of the 
electronically decoupled dye-NiO interface, the interfacial electrostatic is the 
dominant component ruling the relative dye's HOMO metal oxide VB energy levels line-up. 
The dependence of the injection rate on the dye molecular and electronic structure has been investigated by the simple Newns-Anderson
model, showing that
changes in the units constituting the sensitizer (without altering
the conjugation) produces variations both on the energy and on
the spatial localization of the injecting LUMO orbital. This clearly reflects on the energetic alignment
between the dye and TiO$_2$ states and on the coupling between
the interacting species, with inevitable consequences on the
corresponding injection rates. As a final example we have reviewed the effects produced by a shift of the
TiO$_2$ CB edge, simulated by adsorbing an increasing number of
protons  on the semiconductor surface, on the
calculated injection rates. To do that, use was made
of a diabatic-like treatment, which provided us with the
coupling matrix elements between the injecting LUMO orbital and the 
semiconductor CB acceptor orbitals. Addition of protons on the
semiconductor surface, while producing a consistent shift of both
the TiO$_2$ CB and dye LUMO energetic positions, influences the
injection rate mainly through the variations generated in the
electronic coupling.

\section*{Acknowledgements}
MP thanks COHESION and ANR-HELIOSH2 for financial support.

\bibliographystyle{ieeetr}
\bibliography{bibliography.bib}{}

\end{document}